\documentclass[twocolumn,prb,aps,showpacs,amsmath,floatfix,superscriptaddress,amssymb]{revtex4}

\usepackage{graphicx}
\usepackage{dcolumn}
\usepackage{bm}
\usepackage{epsfig}
\usepackage{fancyhdr}
\def\A{\AA$^{-1}\:$}
\def\Am3{\AA$^{-3}\:$}

\def\4he{$^4$He}
\def\3he{$^3$He}

\def\nk{$n({\bf k})\:$}

\begin{document}

\title{Condensate fraction and atomic kinetic energy of liquid ${^3}$He-${^4}$He mixtures}
\author{S. O. Diallo}
\affiliation{Department of Physics and Astronomy, University of Delaware, Newark, DE USA 19716-2570}
\author{J. V. Pearce}
\affiliation{Department of Physics and Astronomy, University of Delaware, Newark, DE USA 19716-2570}
\affiliation{Institut Laue-Langevin, BP 156, 38042 Grenoble, France}
\author{R. T. Azuah}
\affiliation{Department of Materials Science and Engineering, University of Maryland, College Park, MD  20742-2115}
\affiliation{NIST Center for Neutron Research, Gaithersburg, MD 20899-8562 }
\author{F. Albergamo}
\affiliation{European Synchrotron Radiation Facility, BP 220, 38043 Grenoble, France}
\author{H. R. Glyde}
\affiliation{Department of Physics and Astronomy, University of Delaware, Newark, DE USA 19716-2570}

\date{\today}

\begin{abstract}
We present neutron scattering measurements of the momentum distribution of liquid ${^3}$He-${^4}$He mixtures. The experiments were performed at wavevectors $Q$, 26 $\leq$ $Q$ $\leq$ 29 {\AA$^{-1}$}, on the MARI time-of-flight spectrometer at the ISIS Facility, Rutherford Appleton Laboratory, a spallation neutron source. Mixtures with $^3$He concentrations $x$ between 0 and 20\% were investigated both in the superfluid and normal phases. From the data, we extract the Bose-Einstein condensate fraction $n_0$ and the momentum distributions of $^3$He and $^4$He atoms.
We find that $n_0$ increases somewhat above  the pure $^4$He value when $^3$He is added; e.g from $n_0=(7.25\pm0.75)\%$ at $x=$ 0 to $(11\pm3)\%$ at $x=$ 15-20\%. This agrees with  predictions but is less than the only previous measurement. We find a $^4$He kinetic energy $K_4$ for pure $^4$He that agrees with previous determinations. $K_4$ decreases somewhat with increasing $^3$He concentration, less than observed previously and found in early calculations but in agreement with a more recent Monte Carlo calculation.
The $^3$He response is not well reproduced by a Fermi gas momentum distribution, $n(\bf{k})$. Rather an $n(\bf{k})$ having a small step height at the Fermi surface and a substantial high momentum tail characteristic of a strongly interacting Fermi liquid provides a good fit.  This $n(\bf{k})$ is consistent with calculated $n(\bf{k})$. Thus agreement between theory and experiment is obtained comparing $n(\bf{k})$ in contrast to earlier findings based on comparing calculated and observed $^3$He kinetic energies.

\end{abstract}

\maketitle
\setcounter{page}{1}
\section{Introduction \label{sec:intro}}
Liquid ${^3}$He-${^4}$He mixtures are  excellent examples of interacting Bose and Fermi liquids in nature. Their behavior is dominated by quantum-mechanical exchange effects which involve not only identical particles, as in the pure liquids, but also interactions between particles obeying different statistics. As a result, a rich array of macroscopic properties are observed as the $^3$He concentration is increased. Among these properties are a gradual suppression of the $^4$He superfluidity and a finite solubility ($\sim$6.6\%) of $^3$He in $^4$He at absolute zero temperature.

Historically, investigations \cite{Edwards:65,Anderson:66,Ahlers:76,Greywall:79} of this binary mixture involved thermodynamical and hydrodynamical experiments. A central goal was to measure the phase diagram of the mixture and to investigate the elementary excitations as a function of
 $^3$He concentration, $x$. In particular, the $^3$He quasiparticles and the $^4$He phonon-roton (p-r) spectral behavior have drawn much attention.\cite{Surko:73,Rowe:73,Hilton:77,Fak:90}

Fifty years ago, Landau and Pomeranchuk (L-P) proposed a model of the dilute helium mixtures at low temperatures as two interpenetrating fluids of $^3$He and $^4$He atoms for which $^3$He-$^4$He interactions are retained and $^3$He-$^3$He interactions are ignored. In this simple picture, the $^3$He quasiparticle energy is given by $\epsilon_3(k)=\hbar k^2/2m^\ast$ where $m^\ast$ is the $^3$He effective mass in the presence of $^4$He atoms.
At low $k$, the $^3$He excitation spectrum observed in neutron scattering \cite{Hilton:77} agrees well with the L-P spectrum. However, for $k\gtrsim1.5$ \AA$^{-1}$, it departs from the theory and falls below the L-P curve. This observation and the observed small shift in the $^4$He roton energy \cite{Rowe:73,Hilton:77}, $\Delta$, suggest a dynamical level-repulsion between $\epsilon_3(k)$ and $\epsilon_4(k)$ in the vicinity of $\Delta$. A smaller shift in the roton energy was also observed in Raman scattering \cite{Surko:73} for several mixtures concentrations, up to $x=$ 30\%. As a possible explanation, Pitaevskii \cite{Pitaevskii:73} proposed the existence of a small roton-like minimum in $\epsilon_3( k)$. This suggestion found some theoretical support \cite{Stephen:73} but no experimental evidence. 

A fundamental parameter of interest in helium mixtures is the energy-dependent quasiparticle effective mass, $m^\ast$. Yorozu and collaborators \cite{Yorozu:93} reported high accuracy measurements of the $^3$He effective mass in mixtures and found  an $m^\ast$ that strongly depends on pressure but varies little with $x$. Their result was later supported by calculations.\cite{Krotscheck:98}  Recently, the effect of $m^\ast$ on the dynamics of pure liquid $^3$He has been investigated. \cite{Glyde:00a}

  The first neutron scattering measurements at low momentum transfer, $Q$, aimed at probing elementary  excitations in helium mixtures were reported by Rowe {\it et al.} \cite{Rowe:73} for a 5\% mixture at saturated vapor pressure (SVP) and a temperature of 1.6 K and by Hilton {\it et al.} \cite{Hilton:77} for several mixtures at SVP for temperatures in the range 0.6-1.5 K. In these pioniering investigations, the cryogenic capabilities were limited to temperatures close to, or above, the Fermi temperature, $T_F$. In this limit, thermal broadening is important and can mask the quasiparticles energy band. Both Rowe {\it et al.} and Hilton {\it et al.} reported a small shift of the $^4$He p-r curve but disagreed on the sign of the shift.  

F{\aa}k {\it et al.} \cite{Fak:90} were the first to investigate the excitations at temperatures below $T_F$. Their measurement confirmed the small shift in the p-r spectrum observed by Rowe {\it et al.} and by Hilton {\it et al.}. They found no evidence of a roton-like minimum in the $^3$He spectrum. Their experimental results are consistent with the theoretical predictions of Bhatt \cite{Bhatt:78} and of G{\"o}tze {\it et al.}. \cite{Gotze:79}   

Other complementary experimental techniques on helium mixtures include several studies of the elementary excitations by Raman scattering \cite{Woerner:73,Surko:73} and a study of the static structure factor $S(Q)$ by X-ray scattering \cite{Suemitsu:82}. We refer to the review articles by Ahlers \cite{Ahlers:76}, Baym and Pethick \cite{Baym:76} and Glyde and Svensson \cite{Glyde:87} and to the recent book by Dobbs \cite{Dobbs:book} for further details on the subject.

In recent years, inelastic neutron scattering at high $Q$ has been widely used to measure the condensate fraction and atomic kinetic energy of pure liquid $^4$He. \cite{Sears:82,Sears:83,Snow:92,Glyde:book,Azuah:97B,Glyde:00} These high $Q$ measurements are very challenging because of limitations imposed by Instrumental Resolution (IR) and Final-State interactions (FS). A general review of the method and experiments on the VESUVIO neutron spectrometer is presented by Andreani  {\it et al.}. \cite{Andreani:05}
 In experiments involving $^3$He, the large $^3$He neutron absorption cross-section reduces further the scattering intensity and thus the statistical precision of the experimental data. Nevertheless, a number of measurements at high $Q$ have now been performed on pure liquid \cite{Sokol:85,Mook:85,Azuah:95,Azuah:95B,Senesi:03} and solid $^3$He. \cite{Senesi:01} In these measurements, sample cells having advantageous geometries with thin samples or samples at very low $^3$He concentrations are generally used to minimize the effect of absorption. 

In contrast to the pure liquids, there are few reported neutron scattering measurements at high $Q$ of the isotopic helium mixtures. Most report values for the single-particle kinetic energies as a function of $^3$He concentration, $x$, in or close to, the normal phase \cite{Azuah:95,Azuah:95B,Senesi:03}. The only neutron scattering measurement of $n_0$, to date, was reported by Wang and Sokol \cite{Wang:94} in a 10\% mixture where they found an $n_0=18\%\pm3$. 
Several calculations of the Bose-Einstein condensate fraction (BEC), $n_0$, in helium mixtures \cite{Lee:81,Fabrocini:82,Moroni:97,Boronat:90,Boronat:97} predicted an enhancement of $n_0$. 
However, the enhancement predicted by these calculations is significantly less than reported by Wang and Sokol.  

 Theoretical treatments predict that the kinetic energy of both isotopes decreases with increasing concentration. \cite{Lee:81,Boninsegni:95,Boninsegni:97,Boronat:97} Neutron scattering results \cite{Azuah:95,Azuah:95B,Senesi:03} agree both quantitatively and qualitatively with theory for the behavior of $K_4$. In pure liquid $^3$He, there is  a reasonable agreement between recent experiments \cite{Azuah:95,Azuah:95B,Senesi:03} and theory \cite{Whitlock:87} on the value of $K_3$, although earlier observed values were lower than calculated values. \cite{Carlson:85} Mazzanti {\it et al.} \cite{Mazzanti:04} have recently obtained good agreement between theory and experiment by comparing  a calculated dynamic structure factor $J(Q,y)$ directly with the observed $J(Q,y)$.

In dilute helium mixtures, in contrast, there is a significant disagreement in $K_3$. For example, the calculated value of $K_3$ for a 6.6\% mixture is about 19 K [Ref. \onlinecite{Boronat:97}] but the measured value ($\sim$10-12 K) is found to be the same as in the pure liquid $^3$He independent of $x$.
Repeated refinements of the calculation technique using methods as diverse as Monte Carlo diffusion calculations \cite{Boronat:99} to path integral MC techniques \cite{Boninsegni:97} have failed to resolve this disagreement. 

In this paper, we report on deep inelastic neutron scattering measurements aimed at determining $n_0$, $ K_4$, and the $^3$He momentum distribution in $^3$He-$^4$He mixtures. The quantity observed is the dynamic structure factor (DSF) $S(Q,\omega)$.
For a sufficiently large momentum transfer from the neutron to the struck atom, $Q \rightarrow \infty$, the observed DSF reduces to the impulse approximation (IA) \cite{Lovesey:84,Glyde:book},
\begin{equation}
S_{IA}(\mathbf{Q},\omega) = \int d\mathbf{k}n(\mathbf{k})\delta(\omega-\omega_R - 
\mathbf{k.v}_R).
\label{eqn:SQW_ia}
\end{equation}
\noindent where $\omega_{R}=\hbar\mathbf{Q}^2/2m$ and $\mathbf{v}_R=\hbar\mathbf{Q}/m$ are the free atom 
recoil frequency and velocity, respectively. In the IA, $S_{IA}(\mathbf{Q},\omega)$ depends only on a single 
\lq {\it y} scaling'   variable $y=(\omega-\omega_R)/v_R $ and is conveniently expressed as 
\begin{equation}
J_{IA}(y)=v_R S_{IA}(Q,\omega)= \int d\mathbf{k}n(\mathbf{k})\delta(y-k_Q)
\label{eqn:jqy_ia}
\end{equation}
where $ k_Q = \mathbf{k}. \frac{\mathbf {Q}}{\lvert Q\rvert}$. $J_{IA}(y)$ is denoted the longitudinal momentum distribution. 
At finite $Q$, however, the struck atom does not recoil freely but rather interacts with its neighbors. These interactions introduce a Final-State (FS) broadening function $R(Q,y)$ in $J(Q,y)$ (see, e.g. [\onlinecite{Glyde:94,Glyde:book}] for a detailed account of FS effects). Including these final-state interactions (FS), the exact $J(Q,y)$ at high $Q$ is
 \begin{equation} 
J(Q,y)=\int_{-\infty}^{\infty} ds e^{-isy}J(Q,s)=\int_{-\infty}^{\infty}ds e^{iys}J_{IA}(s)R(Q,s)
\label{eqn:jqy_s}
\end{equation}
where $J(Q,s)$ is the intermediate scattering function, $J_{IA}(s)$ is the Fourier transform (FT) of $J_{IA}(y)$ and is the one-body density matrix (OBDM) for displacements along $Q$, and $R(Q,s)$ is the FT of the FS function $R(Q,y)$. 
 The reader is referred to Boronat {\it et al.} \cite{Boronat:93} and Mazzanti {\it et al.} \cite{Mazzanti:01} for a detailed discussion of $S(Q,\omega)$ for the isotopic helium mixtures. 

When there is a condensate, the fraction of atoms with zero momentum transfer, $n_0$, contributes an unbroadened peak to $J_{IA}(y)$ and can be directly extracted from the data. A central goal is to find an appropriate model for the momentum distribution $n(\bf{k})$ including a condensate.  For quantum systems, $n(\bf{k})$ is also not a Gaussian. In earlier work \cite{Glyde:00,Diallo:04}, we found that $n(\bf{k})$ in liquid $^4$He and in solid $^4$He  differs markedly from a Gaussian.     

This report is laid out as follows. The experimental method is presented in section \ref{sec:experimental}, followed by a discussion on the data reduction and analysis methods in section \ref{sec:analysis}.  The results are presented in section \ref{sec:results} followed by  a discussion and conclusion in section \ref{sec:conclusions}. 

\section{Experimental details \label{sec:experimental}}

\subsection{Experiment}
The experiment was performed using the MARI time-of-flight (TOF) spectrometer at the ISIS spallation neutron source at the Rutherford Appleton Laboratory, United Kingdom. MARI is a direct geometry chopper spectrometer in which the time of arrival of a neutron in the detector, measured from when the neutrons leave the moderator, determines its energy loss or gain after scattering from the sample. The momentum transfer depends on both the TOF of the neutron and its scattering angle. More than 900 $^3$He gas detectors provide a coverage of scattering angles between 3$^o$ and 135$^o$ in steps of 0.43$^o$. A large range of momentum and energy transfer can therefore be observed simultaneously. 

\begin{figure}[t]
\centering\epsfig{figure=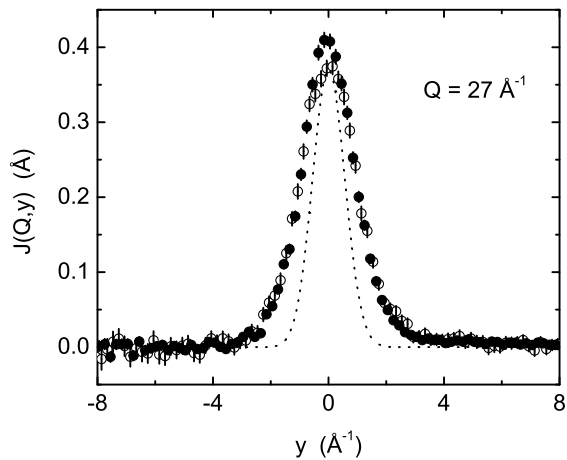,width=3.3in,angle=0}
\caption{\footnotesize MARI Data showing $J(Q,y)$ for pure liquid $^4$He in the normal (open circles) and superfluid phase (closed circles). A higher peak arising from the Bose-Einstein condensate fraction is observed in the superfluid phase. The instrument resolution function is overplotted (dotted line) for comparison.}
\label{fig:high_vs_low}
\end{figure}

An incident neutron energy of 765 meV was selected allowing wavevector transfers up to $Q=$ 30 {\AA $^{-1}$} and energy transfers up to $E=$ 700 meV to be investigated. Measurements of $J(Q,y)$ were made for wavevectors 20 $\leq Q\leq$ 29 {\AA$^{-1}$}. The low energy resolution Fermi chopper was selected to increase count rate in the presence of the highly absorbing $^3$He. This has an energy resolution of approximately 25 meV. The resolution function (in $y$) was determined at each $Q$ as follows. The intrinsic $J(Q,y)$ of pure liquid $^4$He is accurately known from previous measurements on MARI using a high resolution setting.\cite{Glyde:00} Using this $J(Q,y)$ as input, the present broader instrument resolution can be determined directly by reproducing the observed resolution broadened $J(Q,y)$ of pure $^4$He ($x=$ 0). The resolution determined in this way at $Q=$ 27 {\AA $^{-1}$} is shown in Fig. \ref{fig:high_vs_low}.

A special sample cell having a slab geometry, of the type described by Sokol {\it et al.} \cite{Sokol:85} and by F{\aa}k {\it et al.} \cite{Fak:90}, was designed. The sample cell was then placed in the beam and cooled using a $^3$He sorption cryostat. Two separate experiments were performed on MARI under identical experimental conditions. For each set of measurements, a separate background measurement of the empty cell was taken, and the two were found to be very similar, as expected. Data was collected at temperatures of $T$ = 0.4 K, $T$ = 1.3 K and $T$ = 2.5 K which were measured using Ge temperature sensors located at the bottom and top of the sample cell and connected to a Neocera temperature controller. A gas handling system at room temperature was used to prepare mixtures of $^3$He gas to yield liquid  $^3$He concentrations of $x$ = 0, 5, 10, 15, and 20 \%. The pressure above the liquid was maintained at saturated vapour pressure, and was monitored with a pressure transducer by keeping a capillary line open between the gas handling system and the cell.

\subsection{Data Reduction}

Standard procedures were employed to convert the raw neutron scattering data from TOF and scattered intensity to energy transfer and the dynamic structure factor, $S(Q,\omega)$. A summary of these conversions has been given by Andersen {\it et al.}. \cite{Andersen:94b} The data was then converted to the {\textit y} scaling energy transfer variable and to $J(Q,y)=v_{R} S(Q,\omega)$. The data were analyzed in detail for wave vectors 26 $\leq$ $Q$ $\leq$ 29 {\AA$^{-1}$}.

\begin{figure}[t]
\centering\epsfig{figure=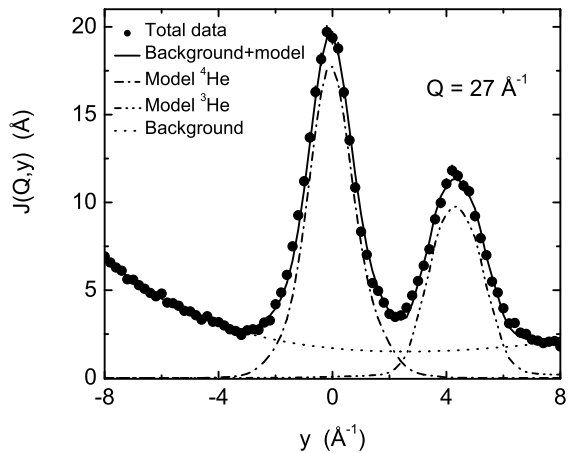,width=3.3in}
\caption{\footnotesize Total observed scattering intensity (solid circles) showing $J(Q,y)$ for a $^3$He-$^4$He mixture at 10\% $^3$He concentration and $T=$ 2.5 K plus background. Dotted line is a separate measurement of the background. The data is presented with the $^4$He peak centered at $y_4=$ 0. The error bars are the size of the circles.}
\label{fig:raw_data}
\end{figure}

The background arising from the empty cell was measured separately and found to be a smooth function of scattering angle and TOF with no microstructure. At a given $Q$ value, the measured background was well represented by a smooth quadratic function in $y$, $ay^2+by+c$, where $a,b$ and $c$ were determined by a least square fit to the measured background. An example of this background is shown in Fig. \ref{fig:raw_data} as a dotted line. This measured background function was subtracted from the scattering from the helium plus cell to obtain the net measured scattering intensity from liquid helium.
 
The analysis consists of representing the net $J(Q,y)$ as a sum of models for the scattering $^4$He and $^3$He. A linear background of variable height and slope was also included to allow for any error in the background determination or shielding of the background by the sample. This linear background was always very small. Additional error in model parameters for $^4$He and $^3$He arising from incorporating the linear background is included in the quoted parameter errors below. The data was {\it y} scaled so that the $^4$He peak is centered at $y=0$ using the $^4$He mass. This means that the $^3$He peak component of the model function must be multiplied by $M=m_3/m_4$ to regain the correct magnitude. An example of the total scattered intensity {\it y} scaled and of models $^4$He and $^3$He plus the measured background is shown in Fig. \ref{fig:raw_data}.

\section{Data Analysis}\label{sec:analysis}

Our goal is to determine the condensate fraction and to get as much information as possible about the $^4$He and $^3$He momentum distributions in the mixture.
 The data is good enough to determine at most two free fitting parameters for each peak.
 To describe the $^4$He peak, we follow the method used for the pure liquid $^4$He. In superfluid $^4$He, we use the convolution approach (CA) in which a FS broadening function $R(Q,y)$ is convoluted with a model $J_{IA}(y)$ as in (\ref{eqn:jqy_s}).
In normal $^4$He, we use the additive approach (AA), described below, and the CA.  The condensate fraction and the width ($\sim$ kinetic energy) of the peak are the free parameters. The FS function and the shape of $n(\bf{k})$ are assumed to be the same as in pure $^4$He. 
The comparison between normal and superfluid liquid $^4$He in Fig. \ref{fig:high_vs_low} shows that the peak height is higher at $y=$ 0 for the superfluid phase due the presence of a condensate and a left-right asymmetry introduced by the condensate term $n_{0}R(Q,y)$ in $J(Q,y)$.
 To describe the $^3$He peak, we construct a model of a Fermi momentum distribution $n(\bf{k})$ and compare the corresponding resolution broadened $J_{IA}(y)$ in (\ref{eqn:jqy_ia}) with the data (no FS effects).
The parameters in the $^3$He $n({\bf k})$ are the step height at the Fermi surface and the length of the high momentum tail.

There is a small overlap of the net $^3$He and $^4$He peaks. Fig. \ref{fig:raw_data} shows that the degree of overlap is very small indeed. The area of overlap of the models is typically 1-2\% of the total area under the two models. 
We found that the parameters in the model fit to one peak were independent of the model used for the second peak within the error quoted. This was tested by using various models to represent the second peak.

\subsection{$^4$He response}
\subsubsection{Convolution Approach}

The convolution approach (CA) is most useful when the momentum distribution $n(\bf{k})$, and thus $J_{IA}(y)$ in (\ref{eqn:jqy_ia}), are narrow in $y$ relative to $R(Q,y)$. Equally, it is useful if $n(\bf{k})$ contains a component, such as a condensate, that is narrow relative $R(Q,y)$. In this case, the observed width of the narrow component in $J(Q,y)$ is set by $R(Q,y)$. In pure liquid $^4$He, the FWHM of $R(Q,y)$ is approximately 1 \AA$^{-1}$. This is broad compared to a condensate component but narrower than the remainder of $n(\bf{k})$.

To determine the condensate fraction, it is convenient to separate the state that is macroscopically occupied from the regular uncondensed states. In a uniform liquid, the natural orbitals are plane wave, momentum states. The orbital containing the condensate is the $k=$ 0 state. This state contributes a term $n_0\delta(\bf{k})$ to $n(\bf{k})$ where $n_0=N_0/N$ is the fraction of particles in the condensate. The regular, uncondensed states are the $k\neq$ 0 states and we denote their contribution to $n(\bf{k})$ as $n^{\ast}(\bf{k})$. In an interacting Bose liquid, Bosons can scatter into and out of the condensate from the $k\neq$ 0 states. This leads to a coupling
 between the condensate and the $k\neq$ 0 states and a term \cite{Griffin:book,Glyde:94,Glyde:00}

\begin{equation}
n_{0} f({\bf k}) =
\left[
\frac{n_{0}mc}{2\hbar\left(2\pi^{3}n\right)}
\frac{1}{|{\bf k}|}
\coth\left(\frac{c\hbar |{\bf k}|}{2k_{B}T}\right)
\right]
e^{-k^{2}/(2k_{c}^{2})}
\label{eqn:f_of_k}
\end{equation}
\noindent in $n(\bf{k})$. 
The coupling is strongest for the low $k$ states and the expression in the square bracket of (\ref{eqn:f_of_k}) is derived for and valid for the low $k$ limit. We have multiplied it by a Gaussian to cut off $n_0f(\bf{k})$ at higher $k$ as done for pure \cite{Glyde:00} $^4$He with $k_c$ fixed at $k_c=0.5$ {\AA$^{-1}$}.   This term is highly localized around $k=$ 0 and when Final State broadening is included $n_0f(\bf{k})$ cannot be distinguished experimentally from $n_0\delta(\bf{k})$. The model $n(\bf{k})$ we use for the $^4$He component therefore has three terms \cite{Sears:82},
\begin{equation}
n({\bf k})=n_{0}[\delta ({\bf k})+f({\bf k})] + A_{1}n^{\ast} ({\bf k}).
\label{eqn:mom_distribution}
\end{equation}
\noindent $A_{1}$ is a constant chosen by normalization $\int d{\bf k} n({\bf k})=$ 1.

To implement this model, we Fourier transform $n(\bf{k})$ to obtain the corresponding One-Body-Density-Matrix (OBDM), $n(\bf r)$=$\int d{\bf k} e^{i\bf{k.r}} n({\bf k})$. 
$J_{IA}(s)$ is the OBDM for displacements ${\bf r}=s{\bf\hat Q}$ parallel to the scattering wavevector $\bf{Q}$. $J_{IA}(s)$ corresponding to (\ref{eqn:mom_distribution}) is obtained by Fourier transforming (\ref{eqn:mom_distribution}) to obtain $n({\bf r})$ and using $J_{IA}(s)=n(s)$ giving,
\begin{equation}
J_{IA}(s)=n(s)=n_{0}[1+f(s)] + A_{1}n^{\ast}(s)
\label{eqn:OBDM}
\end{equation}

The term $[1+f(s)]$ is long range in $s$. The $n^{\ast}(s)$ is short range in $s$ and we represent it by \cite{Glyde:00}

\begin{equation}
n^{\ast}(s)=\exp\left[-\frac{\bar\alpha_2}{2!}s^2+\frac{\bar\alpha_4}{4!}s^4-\frac{\bar\alpha_6}{6!}s^6\right]
\label{eqn:ns_cumulants}
\end{equation}

\noindent where $\bar\alpha_2$, $\bar\alpha_4$ and $\bar\alpha_6$ are parameters (cumulants).
This representation is useful when $n^{\ast}(s)$ is at least approximately a Gaussian, plus small corrections.
 In going from (\ref{eqn:mom_distribution}) to (\ref{eqn:OBDM}), we actually need to transform $f(\bf{k})$ only, and once only since $f({\bf k})$ has no free parameters. Given $f(s)$, our model may be viewed as a model for $J_{IA}(s)$ in (\ref{eqn:OBDM}).

The model $J_{IA}(s)$ is multiplied by the FS function $R(Q,s)$ and the product is Fourier transformed to obtain $J(Q,y)$ as in (\ref{eqn:jqy_s}). In the present fit to the $^4$He peak, we used the pure $^4$He FS function at all $^3$He concentrations,
\begin{equation}
 R(Q,s)=\exp\left[-\frac{i\bar\beta_3}{3!}s^3-\frac{i\bar\beta_5}{5!}s^5-\frac{\bar\beta_6}{6!}s^6\right]
\label{eqn:fs_cumulants}
\end{equation}
with parameters $\bar\beta_3=\bar a_3/\lambda Q$, $\bar\beta_5=\bar a_{52}/(\lambda Q)^3$ and $\bar\beta_6=\bar a_{64}/(\lambda Q)^2$ with $\bar a_3, \bar a_{52}$ and $\bar a_{64}$ set at their pure $^4$He values.\cite{Glyde:00} 
We also set $\bar\alpha_4$ and $\bar\alpha_6$ in $n^\ast(s)$ at their pure $^4$He values.\cite{Glyde:00} The $J(Q,y)$ therefore has only two free parameters, $\bar\alpha_2$ and $n_0$.

The above procedures for determining the shape of the momentum distribution of pure liquid $^4$He, and extracting the condensate fraction are well established, and have been the subject of intense activity  over the years.\cite{Sears:82,Griffin:book,Glyde:94,Glyde:00}

\subsubsection{Additive Approach}

When the momentum distribution is broad, FS effects are relatively less important. In this case, the exponential in $R(Q,s)$ of (\ref{eqn:fs_cumulants}) can be expanded and FS effects retained as additive corrections to the IA. Similarly, if the deviations of $n^{\ast}(s)$  from a Gaussian  are not too large, $n^{\ast}(s)$ in (\ref{eqn:ns_cumulants}) can be expanded and deviations from a Gaussian $n^{\ast}(s)$ retained as additive corrections. This expansion of $n^\ast(s)$ is the same as the Gauss-Hermite expansion introduced by Sears.\cite{Sears:83} 
These expansions lead to the Additive Approach (AA), in which $J(Q,s)$ is represented as a Gaussian IA, $\bar J_{IA}(s)$, plus corrections for deviations of $n^{\ast}(s)$ from a Gaussian and for FS effects as \cite{Glyde:book,Andersen:97},
\begin{equation}
J(Q,s)=\bar J_{IA}(s) + J_{1}(Q,s) + J_{2}(Q,s) - \dots
\label{eqn:aa}
\end{equation}
\noindent where
\begin{eqnarray}
\nonumber \bar J_{IA}(s)=\left(2\pi \bar \alpha_{2}\right)^{-1/2}
e^{-s^2/2\bar \alpha_{2}}\\
\nonumber J_{1}(s)=\frac{i}{3!}\bar\mu_{3}s^3 \bar J_{IA}(s)\\
J_{2}(s)=\frac{1}{4!}\bar\mu_{4}s^4\bar J_{IA}(s)
\label{eqn:aa_components}
\end{eqnarray}
\noindent and $\bar\mu_3=\bar a_3/\lambda Q$ and $\bar\mu_4=\bar\alpha_4+\bar a_4/(\lambda Q)^2$. The $\bar\alpha_4$ is the leading deviation from a Gaussian and $\bar a_3$ and $\bar a_4$ are FS terms. In fitting this function to the $^4$He peak in normal mixtures, we fixed $\bar a_3$ and $\bar\alpha_4$ at their pure $^4$He values ($\bar a_4=0$). The $J(Q,y)$ then has only one free parameter, $\bar\alpha_2$.

\subsection{$^3$He Response}

 The $^3$He momentum distribution is not well approximated by a Gaussian. Thus we do not expect the fit of a Gaussian or the AA to the $^3$He peak to reveal the physics well.  To proceed, we constructed a simple model of the $^3$He $n(\bf{k})$ consisting of a step of height $Z$ at the Fermi surface plus an exponential high momentum tail. We calculate the $J_{IA}(y)$ for this $n(\bf k)$ using (\ref{eqn:jqy_ia}) and fit the convoluted $J_{IA}(y)$ directly to the data. 

 Specifically, the model $n(\bf k)$ is,
\begin{equation}
n(\bf{k})=
\begin{cases}
\frac{H}{V_F}& \text{$\frac{k}{k_F}< 1$} \\
\frac{H_F}{V_F}e^{-\eta\left(\frac{k}{k_F}-1\right)}& \text{$\frac{k}{k_F}>1$}
\end{cases}
\label{eqn:fermi_nk}
\end{equation}
\noindent where $k_F$ is the Fermi momentum and $V_F=\frac{4\pi}{3}k_F^3$ the volume of the Fermi sphere.
Assuming that the volume occupied by $^3$He in the mixtures exceeds that occupied by $^4$He by about 28\%, as suggested by the dielectric measurements of Edwards et al. \cite{Edwards:69}, the $k_F$ can be approximated using the expression $k_F=({3\pi^2\rho x})^{1/3}$ where $\rho$ is the average density of the mixture at SVP and $x$ the $^3$He concentration. The $k_F$ values are listed in Table \ref{tbl:I}.
 \begin{figure}[t]
\centering\epsfig{width=3.3in,figure=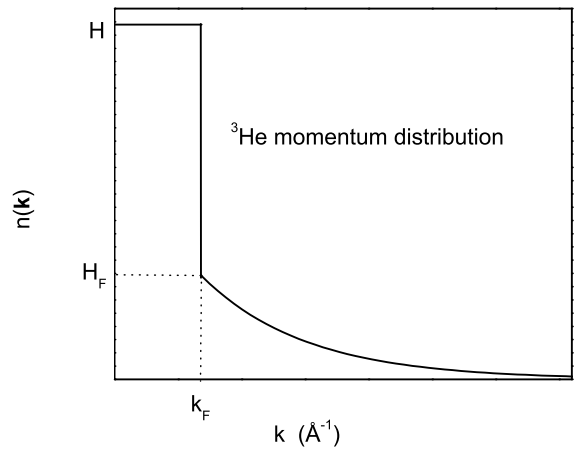}
\caption{\footnotesize Schematic diagram of the Fermi model momentum distribution used in this analysis. The parameters are defined in the text.}
\label{fig:modelnk}
\end{figure}
\begin{table}[b]
\caption{\footnotesize Fermi momentum $k_F(x)=({3\pi^2\rho x})^{1/3}$  as a function of the $^3$He concentration $x$ where $\rho$ is the average mixture density at SVP. 
}
\begin{ruledtabular}
\begin{tabular}{c| c c c c c}
 $x$ (\%)  & 5 & 10 & 15 &20 & 100\\           
\hline
$k_F$ (\AA$^{-1}$)&0.32&0.40&0.45 &0.50&0.79 
\end{tabular}
\label{tbl:I}
\end{ruledtabular}
\end{table}
The model has three parameters, $H$, $H_F$ and $\eta$ where $\eta$ determines the length of the high momentum tail and $Z=H-H_F$ (See Fig. \ref{fig:modelnk}). Normalization of $n(\bf k)$, $\int d{\bf k} n({\bf k})=1$, gives  
\begin{equation}
H +\frac{3H_F}{\eta}\left(\frac{2}{\eta^2}+\frac{2}{\eta}+1\right)=1
\label{eqn:normalization}
\end{equation}
\noindent which can be used to eliminate one parameter, say $H$. 
Similar models of $n(\bf{k})$ are discussed by Carlson {\it et al.} \cite{Carlson:85} and used by Azuah {\it et al.} \cite{Azuah:95B} and Mazzanti {\it et al.}\cite{Mazzanti:04} for pure liquid $^3$He.
The IA given by (\ref{eqn:jqy_ia})
corresponding to (\ref{eqn:fermi_nk}) is,
\begin{equation}
J_{IA}(y')=
\begin{cases}
       \frac{3}{4k_F}\left(H(1-y'^2)+\frac{2H_F}{\eta}(1+\frac{1}{\eta})\right) & y'< 1  \\
       \frac{3}{4k_F}\left(\frac{2H_F}{\eta}(\frac{1}{\eta}+y')e^{-\eta(y'-1)}\right) &  y'>1
\end{cases}
\label{eqn:jia_he3}
\end{equation}
\noindent where $y'=y/k_F$. 

If the liquid temperature $T$ is near or above the Fermi temperature,  $T_F=\frac{\hbar^2}{2m^\ast k_B}k_F^2$, there may be some thermal broadening of the Fermi liquid $n(\bf k)$.\cite{Dobbs:book} This is certainly the case for a Fermi gas for which $n(\bf k)$ is a step function ($Z=1$) at $T=0$ K. However, in the liquid model above, interaction has already reduced the magnitude of the step at $k_F$ ($Z<1$). Thus thermal broadening  may be relatively less important in a strongly interacting liquid. This is discussed more fully in section \ref{sec:results}.

 To estimate $T_F$, we note that the effective mass $m^{\ast}$ 
depends strongly on the pressure but little on the $^3$He concentration.\cite{Fak:90} We take $m^{\ast}=2.3m$ as found by theoretical calculations \cite{Boninsegni:95,Krotscheck:98}, independent of concentration. In the range $x=$ 5-20 \%, the corresponding $T_F$ is 0.3-0.8 K. This is significantly less than the normal liquid temperature $T=$ 2.5 K and comparable to the superfluid temperature $T=$ 0.4 K used here. Our $n({\bf k})$ extracted from data will represent $n({\bf k})$ at these temperatures. However, we did not observe any difference between $n({\bf k})$ at $T=$ 0.4 K and 2.5 K.
This suggests that the rounding of the small step $Z$ at $k_F$ by FS effects and the instrument resolution function is comparable or greater than thermal broadening.

\section{Results}\label{sec:results}

In this section, we present our neutron scattering data and the results for the $^4$He atom kinetic energy, the $^3$He momentum distribution and the $^4$He condensate fraction obtained from analyzing the data. Since the absorption cross section of $^3$He is so large, the net scattered neutron intensity from $^3$He-$^4$He mixtures is weak. For this reason, a broad energy resolution setting was used on the MARI spectrometer, as shown in Fig. \ref{fig:high_vs_low}. Also, the statistical precision of the data is not high. As a result, we were able to determine uniquely only one or two parameters in model fits to the data. Specifically, we found that the parameter $n_0$ that provides the condensate fraction and the parameter $\bar\alpha_2$ that sets the kinetic energy, $K_4=3\hbar^2\bar\alpha_2/2m$, were correlated. For this reason, we begin with  normal $^4$He where $n_0=0$.

\subsection{Normal phase}
\subsubsection{$^4$He kinetic energy}
\begin{figure}[t]
\centering\epsfig{figure=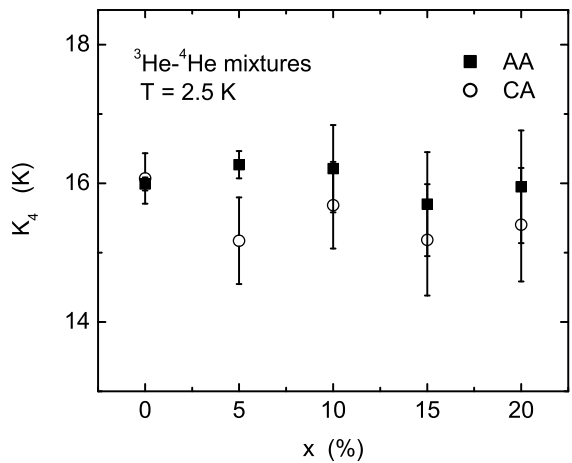,width=3.3in}
\caption{\footnotesize $^4$He kinetic energy [$K_4$] at $T=$ 2.5 K as a function of $^3$He concentration, $x$. The closed squares are the $^4$He kinetic energies obtained from the Additive Approach (AA) and the open circles obtained from the  Convolution Approach (CA).\cite{Glyde:book}}
\label{fig:xdep_k4}
\end{figure}

 The kinetic energy $K_4$ of $^4$He in the normal liquid at $T=2.5$ K is shown in Fig. \ref{fig:xdep_k4} as a function of $^3$He concentration. $K_4$ is determined from the $^4$He peak using two fitting procedures, the convolution approach (CA) and the additive approach (AA). In the CA, we use the Final State function that was determined \cite{Glyde:00} previously for pure $^4$He  unchanged at all $^3$He concentrations. The only free parameter in the present fit was $\bar\alpha_2$. Similarly, in the AA the FS parameters were set at their pure $^4$He values \cite{Glyde:00} and only $\bar\alpha_2$ was free. The parameters held fixed in the CA are summarized in Table \ref{tbl:II}.
\begin{table}[b]
\caption{\footnotesize Fitting variables used in the CA. In the normal phase, only $\alpha_2$ is determined from a fit to data  and in the superfluid phase only $n_0$ is determined. The values of $k_c$, $\beta_n$ and $\alpha_n$ for $n\ge3$ were all kept fixed at their values in the pure $^4$He case, as evaluated by the precise measurements of Glyde {\it et al.}.\cite{Glyde:00}}
\vspace{0.25cm}
\begin{ruledtabular}
\begin{tabular}{c| c c c c}
 $^4$He peak   &$\alpha_2$&$n_0$& $\beta_{n\ge3}$&$\alpha_{n\ge3}$\\           
parameters     &  (\AA$^{-2}$)&(\%)&(\AA$^{-n}$)& (\AA$^{-n}$)\\
\hline
$T=2.5$ K    & Free & 0& Fixed& Fixed\\
\hline
$T=0.4$ K& Fixed&Free &Fixed&Fixed 
\end{tabular}
\label{tbl:II}
\end{ruledtabular}
\end{table}

Fig. \ref{fig:xdep_k4} shows that the CA and AA procedures give identical kinetic energies for pure $^4$He ($x=0$). This value agrees with our previous\cite{Glyde:00} $^4$He determination $K_4=$16.3$\pm$0.3 K  and with the value found by Senesi {\it et al.}.\cite{Senesi:03} 
As the $^3$He concentration is increased, the two methods continue to agree within error but the CA method gives a marginally lower $K_4$. The $K_4$ decreases only very little with increasing $x$. Azuah {\it et al.} \cite{Azuah:95} and Senesi {\it et al.} \cite{Senesi:03} find that $K_4$ decreases more sharply with increasing $x$. A decrease is anticipated since the density decreases as $x$ increases and in pure fluids $K_4$ decreases as density decreases. When we fit a Gaussian to the $^4$He peak, we find a $K_4$ that is independent of $x$.  We return to this point in section \ref{sec:resultsC} and compare with calculations in section \ref{sec:conclusions}.

\begin{figure}[t]
\noindent
\centering\epsfig{figure=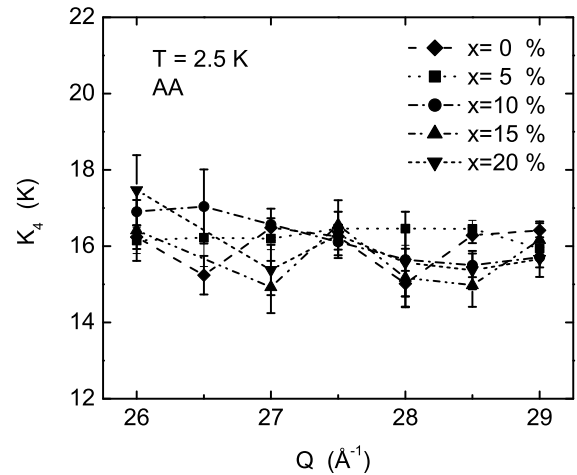,width=3.3in,angle=0}
\centering\epsfig{figure=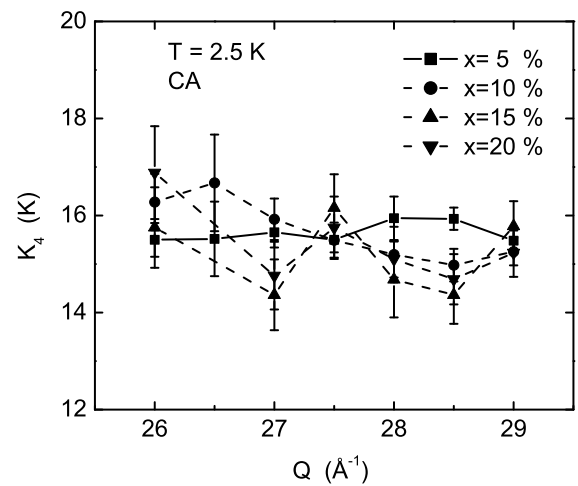,width=3.3in,angle=0}
\caption{\footnotesize Top: Additive Approach (AA) fit to data in the normal phase ($T=$ 2.5 K) at several $^3$He concentrations showing the $Q$ dependence of $ K_4$. Bottom: Convolution Approach (CA) fit to data in the normal phase.}
\label{fig:qdep_k4}
\end{figure}

The $K_4$ values in Fig. \ref{fig:xdep_k4} are averages obtained from fits to data at specific $Q$ values. The variation of $K_4$ with $Q$ is shown in Fig. \ref{fig:qdep_k4}. The aim is to display the statistical precision of the data.  

\subsubsection{$^3$He momentum distribution}
In this section, our goal is to learn as much as possible about the $^3$He momentum distribution. We do this by fitting the $J_{IA}(y)$ (\ref{eqn:jia_he3}) obtained by substituting the model $n(\bf{k})$ given by (\ref{eqn:fermi_nk}) and shown in Fig. \ref{fig:modelnk}, into (\ref{eqn:jqy_ia}). The model $n({\bf k})$ has two parameters, $H_F$ giving the height of the high momentum tail at $k_F$ and $\eta$ which sets the length of the tail. $H$ is determined by normalization.

The data in $^3$He peak region was precise enough to determine one parameter well with an estimate of a second parameter. Since the tail of $n({\bf k})$ is of specific interest, we also set the tail parameter at specific values e.g. no tail ($H=$ 1), tail that corresponds to the calculated $n({\bf k})$ \cite{Boronat:97}  ($\eta=$ 0.8) and fitted for $H_F$ to see how good a fit could be obtained. We subsequently obtained best fit values of $\eta$. Once, $\eta$ is determined, the model kinetic energy is obtained as $K_3 = \int d{\bf k} k^2 n({\bf k})$ giving,
\begin{equation}
 K_3 = \frac{3\hbar^2}{2m_3}\frac{k_{F}^2}{5}\left( H+\frac{5H_F}{\eta}(\frac{4!}{\eta^4} +\frac{4!}{\eta^3}
+\frac{12}{\eta^2}+\frac{4}{\eta} +1) \right)
\label{eqn:k3}
\end{equation}
\noindent where $\frac{\hbar^2}{2m_3}=$ 8.08 K \AA$^{2}$.
For typical values of $\eta$, more than 95\% of $K_3$ arises from the tail in $n({\bf k})$. Also, in (\ref{eqn:k3}), $K_3$ is very sensitive to $\eta$. Thus, unless $\eta$ can be accurately determined, $K_3$ is not a good single parameter to characterize the data.

\begin{figure*}
\begin{minipage}{6.2in}{
\includegraphics[width=0.49\linewidth]{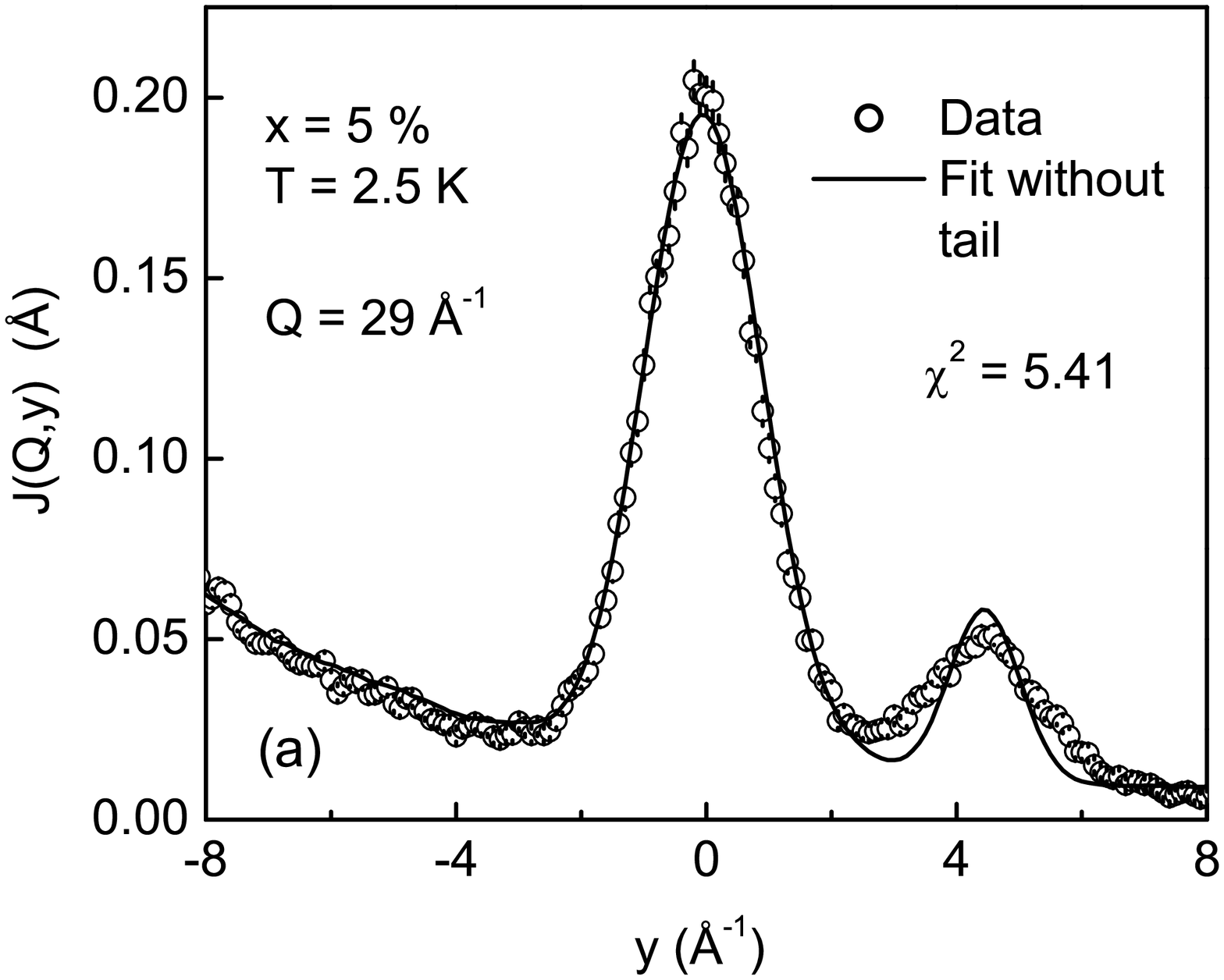}
\includegraphics[width=0.49\linewidth]{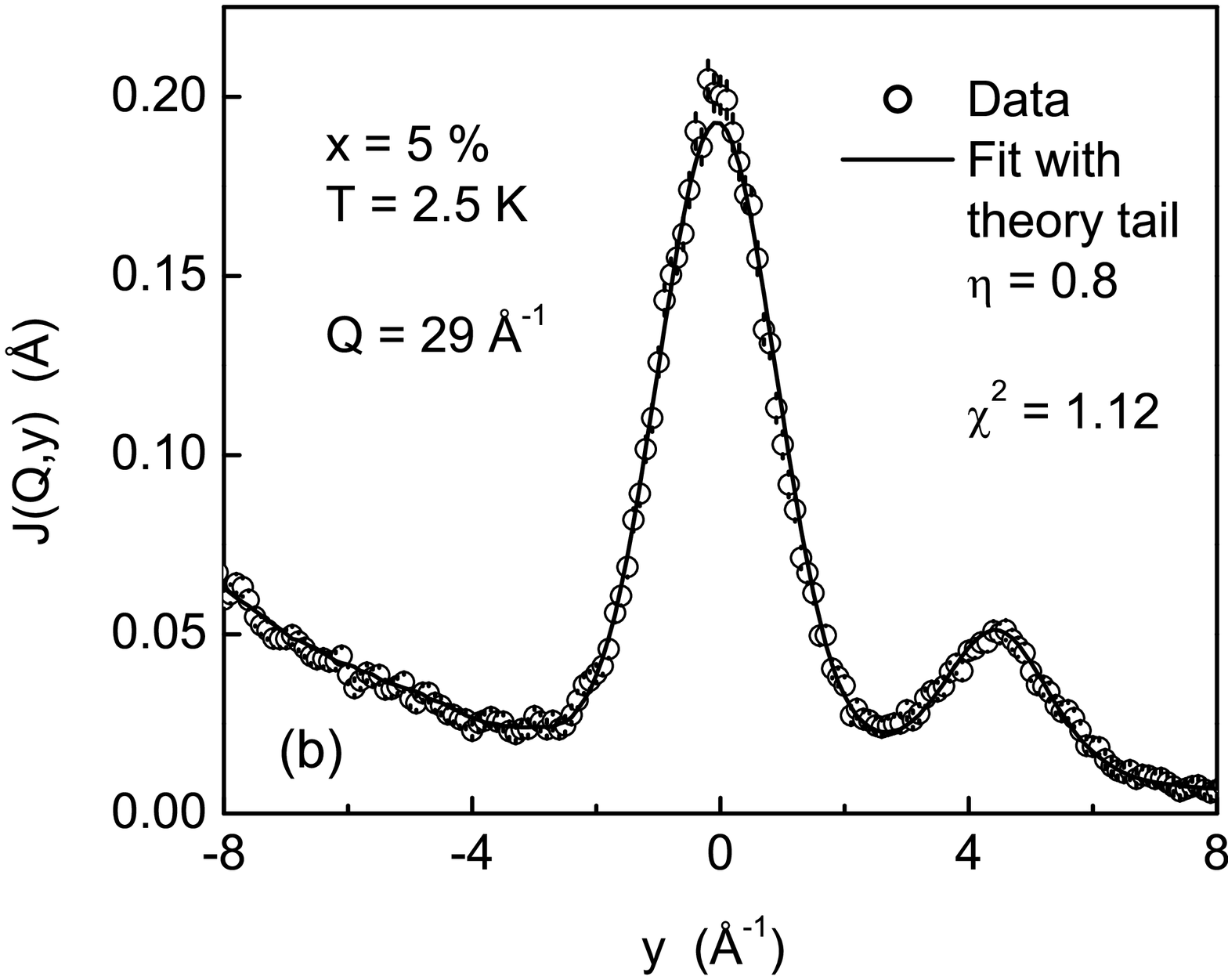}}
\caption{\footnotesize Fits to the $^3$He peak (solid lines) in the normal helium mixture data at $x=$ 5\% and $Q =$ 29 \AA$^{-1}$. Data are open circles with error bars. The error bars are the size of the circles except where shown. The left side shows a fit assuming the $^3$He $n({\bf k})$ is a Fermi step function (i.e $n(\bf{k})$  has no high momentum tail ($H=$1)) and the right side shows a fit to data with the model $n({\bf k})$ shown in Fig. \ref{fig:modelnk} that reproduces the calculated tail of Ref. [\onlinecite{Boronat:97}]. The fit with no tail (left) is poor while the fit incorporating the calculated tail (right) is good.}
\label{fig:jqyfits0}
\end{minipage}
\end{figure*}

\begin{figure*}
\begin{minipage}{6.2in}{
\includegraphics[width=0.49\linewidth]{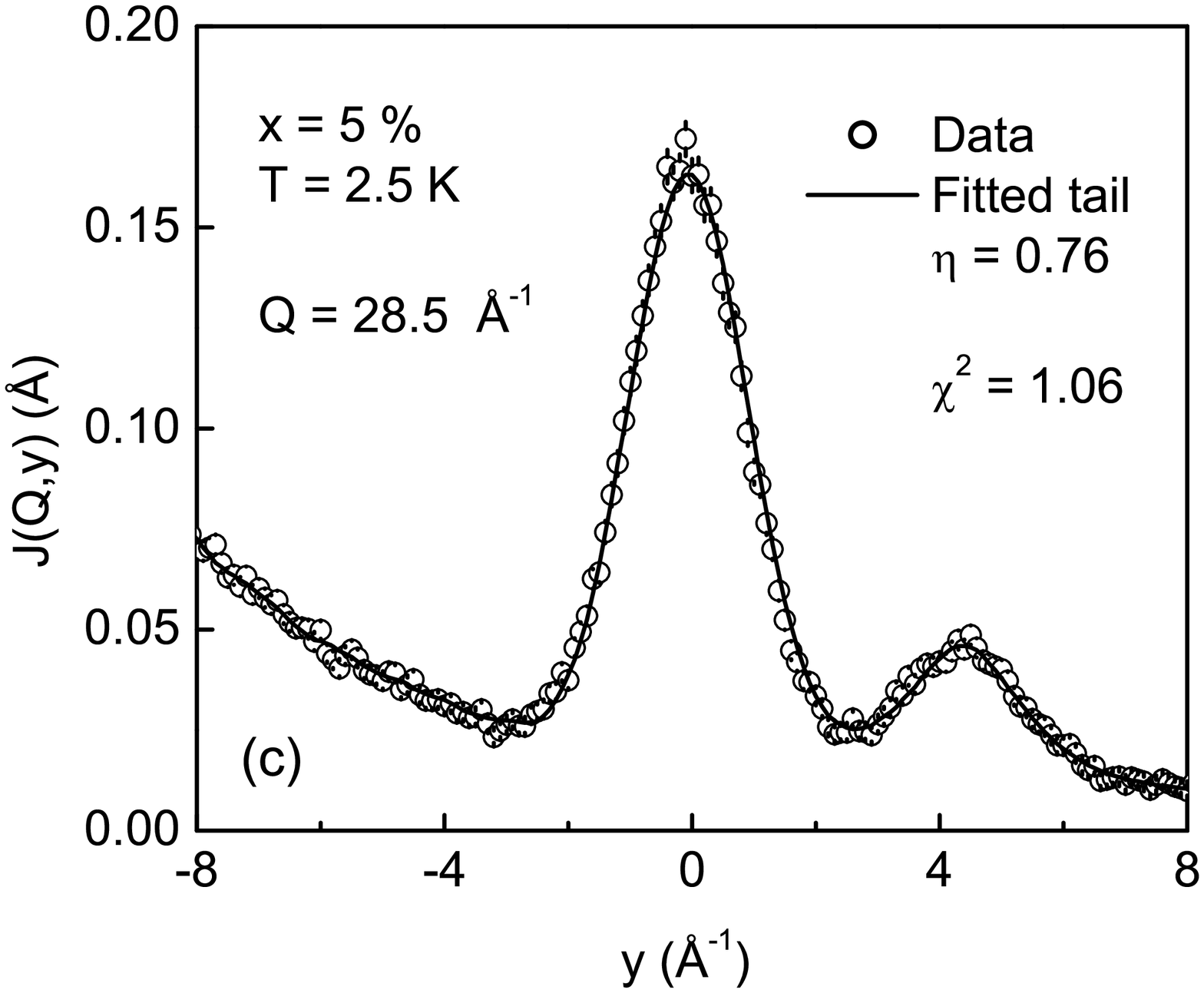}
\includegraphics[width=0.49\linewidth]{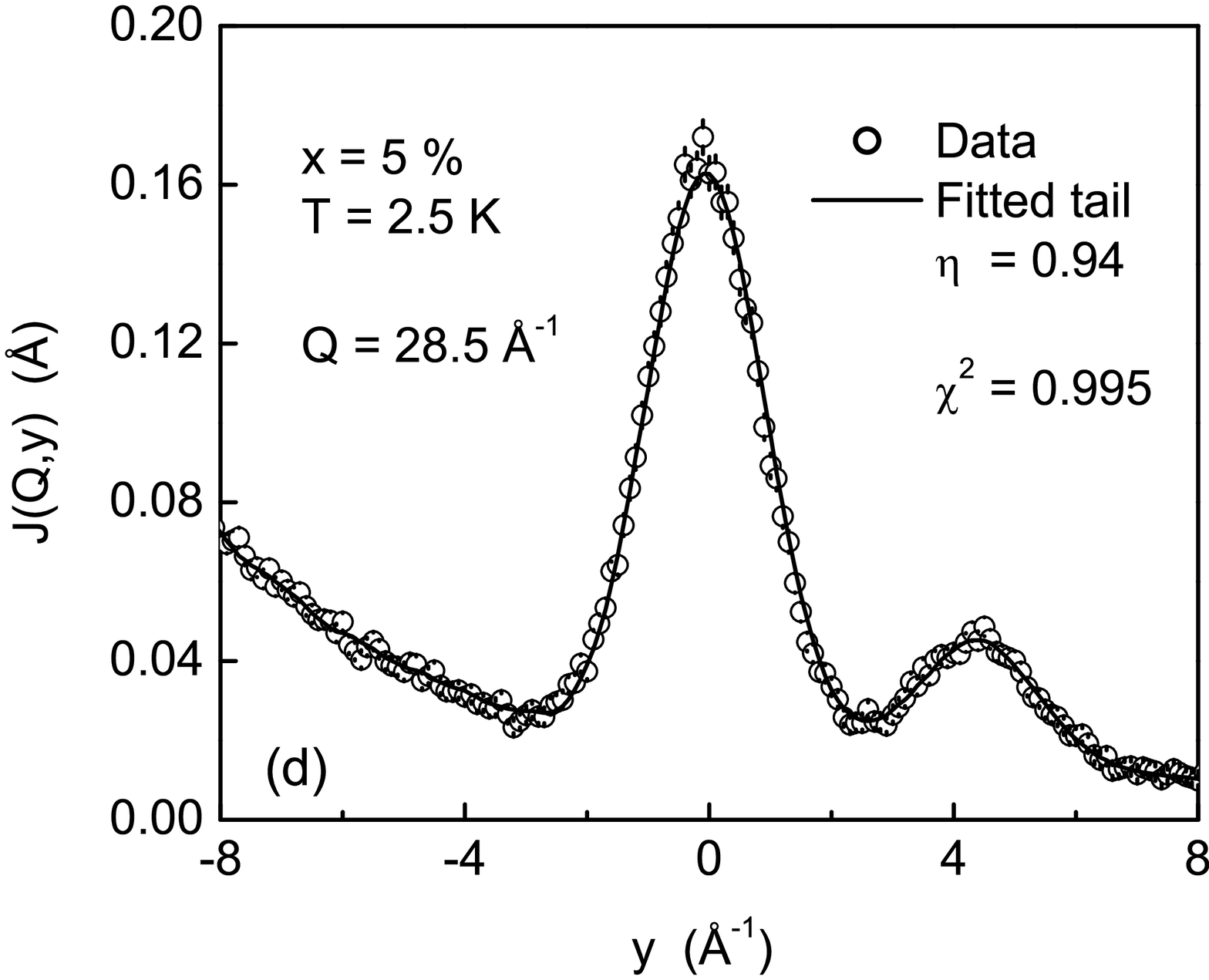}}
\caption{\footnotesize Fits to the $^3$He peak (solid lines) in the normal helium mixture data (open circles) at $x=$5\% and $Q =$ 28.5 \A. The error bars are the size of the circles except where shown. The graphs show that it is possible to get good fits to the same data with two different sets of parameters. This illustrates the precision in which the $\eta$ parameter can be determined. The calculated tail from Ref. [\onlinecite{Boronat:97}] corresponds to $\eta=$ 0.8.}
\label{fig:jqyfits}

\end{minipage}
\end{figure*}
 
Fig. \ref{fig:jqyfits0} shows fits to the $^3$He peak with no high momentum tail 
($Z=H=1, H_F=0$) and with the tail parameters set at $\eta=$ 0.8, the value obtained in a fit to the high $k$ portion of the tail calculated by Boronat {\it et al.}. Clearly a Fermi step function with no tail cannot reproduce the data. In contrast, an $n(\bf{k})$ with a high energy tail that reproduces the calculated tail (see Fig. \ref{fig:jqyfits0}) fits the data. In this sense, the calculated $n(\bf{k})$ and the data are entirely consistent. There is no disagreement between theory and experiment.

\begin{figure}
\centering\epsfig{figure=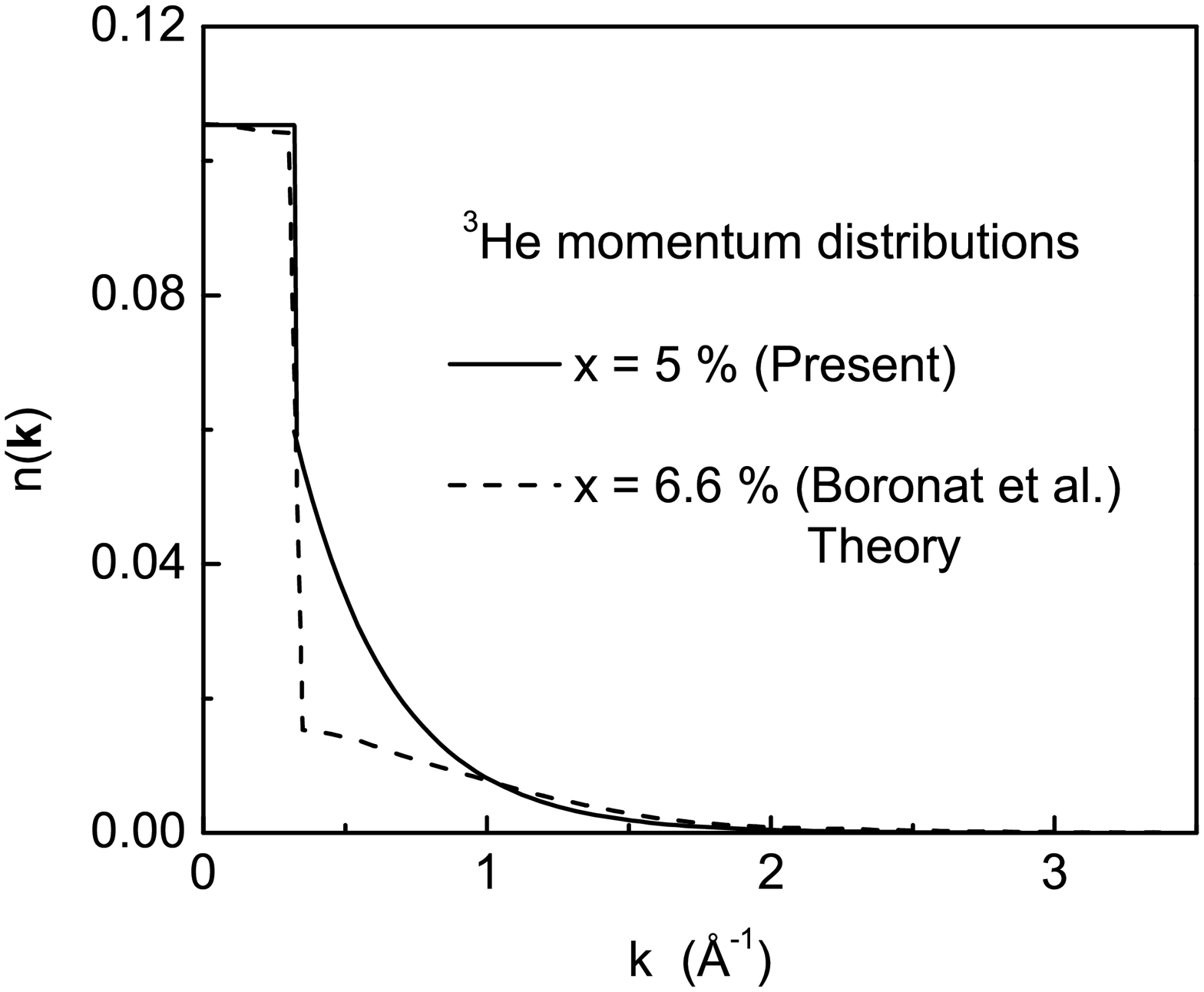,width=3.3in}
\caption{\footnotesize Comparison of the observed $^3$He atom momentum distribution $n(\bf{k})$ (solid line) at $x=$5\% with the calculated $n(\bf{k})$ from 
Ref. [\onlinecite{Boronat:97}] at $x=$ 6.6\% (dotted line).}
\label{fig:nkcompared}
\end{figure}

We determined an observed value of $\eta$.
The best fit value of $\eta$ obtained varied somewhat with the data set (e.g $Q$ value) considered reflecting the statistical precision of the data. Fig. \ref{fig:jqyfits} shows fits for $\eta=$ 0.76 and 0.94 which provide equally good fits at different $Q$'s. The data is precise enough to determine that $\eta$ lies in the range 0.7 $\lesssim \eta \lesssim$ 1.0, i.e $\eta=$ 0.85$\pm$0.15 and $Z=H-H_F=0.05\pm{^{0.14}_{0.01}}$. 
This correspond to a very strongly interacting Fermi liquid. Fig. \ref{fig:nkcompared} shows a comparison between our observed $n({\bf k})$ for a 5\% mixture at $T=2.5$ K and a calculated \cite{Boronat:97} $n({\bf k})$ at 6.6\%.
\subsection{Superfluid phase} 
As noted, we found that the parameter $\bar\alpha_2$ and the condensate fraction $n_0$ were correlated. For example, increasing $n_0$ and decreasing $\bar\alpha_2$ both lead to a narrower peak that cannot be distinguished within the present precision of the data. In pure $^4$He, $\bar\alpha_2$ was found to be the same in normal ($T=$ 2.5 K) and superfluid ($T=$ 0.5 K) $^4$He. Thus in the mixtures we set $\bar\alpha_2$ in the superfluid phase at the normal phase value determined above at each $x$.

\subsubsection{Condensate fraction}
\begin{figure}[t]
\centering\epsfig{figure=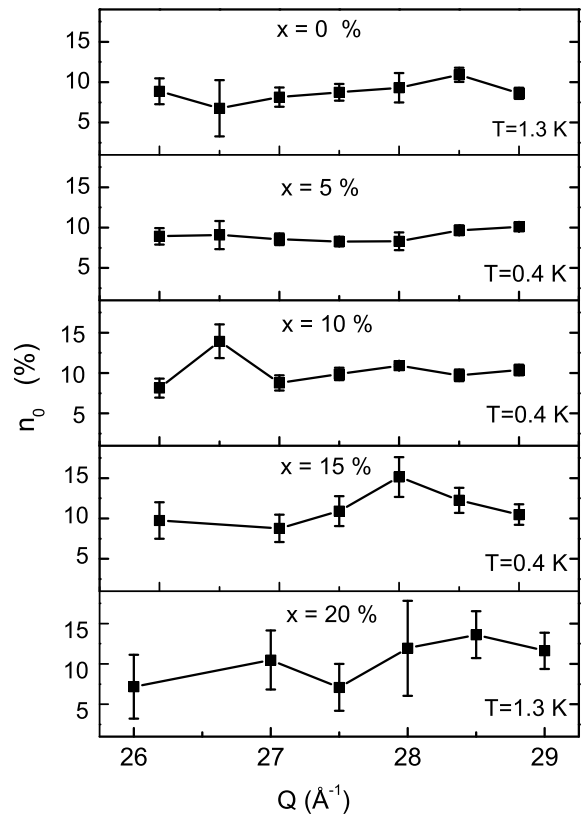,width=3.3in}
\caption{\footnotesize The $^4$He condensate fraction (closed squares), $n_0$, as a function of momentum transfer $Q$. The error bars represent statistical errors obtained from fits to data. The solid lines are guides to the eye.}
\label{fig:qdep_n0_lowt}
\end{figure}

\begin{figure}[t]
\centering\epsfig{figure=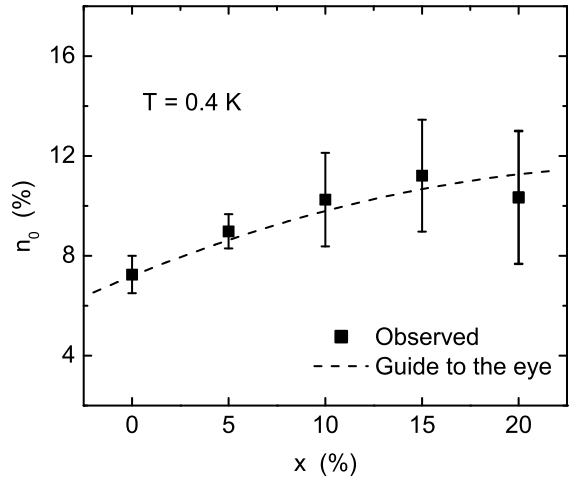,width=3.3in,angle=0}
\caption{\footnotesize The $^4$He condensate fraction (closed squares), $n_0$, as a function of $^3$He concentration. The error bars are standard deviations obtained from a linear least-squares fit to the $Q$ dependence of $n_0$. The condensate fraction increases with increasing $x$ from 7.25$\pm$ 0.75\% at $x=$ 0 to 11$\pm$3 at $x=$ 15-20\%. The dashed line is a guide to the eye.}
\label{fig:xdep_n0_lowt}
\end{figure}

To determine $n_0(x)$, we first perform fits of the CA to data for each $Q$ at a given concentration $x$. The average value over all $Q$ gives the corresponding $n_0(x)$. The variation of the condensate fraction with $Q$ is shown in Fig. \ref{fig:qdep_n0_lowt}. The variation reflects the statistical precision of the data.  Fig. \ref{fig:xdep_n0_lowt} shows the resulting $x$ dependence.
The helium mixtures data at $x=20\%$ was taken at a slightly higher temperature ($T$ = 1.3 K) than the lower concentrations data. To be able to make a consistent comparison of the results on $n_0$, we made a temperature correction \cite{Glyde:00} to $n_0$ at $x=20\%$ using  $n_0(T)=n_{0}[1-(\frac{T}{T_{\lambda}})^{\gamma}]$ where $T_\lambda \sim 2.17$ and $\gamma=5.5$ to get its equivalent value at $T=0.4$ K. The corrected $n_0$ is shown in Fig. \ref{fig:xdep_n0_lowt} along with the results at lower $^3$He concentrations. 
We find that $n_0(x)$ increases slightly above the pure $^4$He value with increasing $x$. Fig. \ref{fig:xdep_n0_lowt} represents our final results for $n_0$ which are discussed in section \ref{sec:conclusions}.

\subsubsection{$^4$He kinetic energy}
\begin{figure}[t]
\centering\epsfig{figure=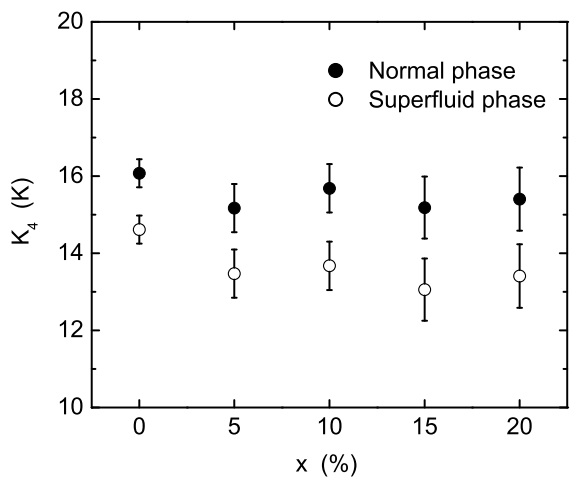,width=3.3in}
\caption{Kinetic energy of $^4$He atoms, $K_4$, as a function of concentration. The superfluid $K_4$ were obtained from the normal $K_4$ assuming the condensate does not contribute to the kinetic energy.}
\label{fig:K4_superfluid}
\end{figure}

The $^4$He kinetic energy in the superfluid phase can be obtained using the form of the momentum distribution in (\ref{eqn:mom_distribution}) and assuming that the parameter $\bar\alpha_2$ is the same in both normal and superfluid phases. 
Normalizing the $^4$He model momentum distribution, as discussed above, leads to $n_0[1+I_f]+A_1=1$ with $I_f\sim$ 0.25. This result can be obtained from (\ref{eqn:OBDM}) as $n(s=0)$ where $I_f=f(s=0)$ and $n^{*}(s=0)=1$. Since $n_0$  is a function of $x$, the normalizing constant $A_1$ is also a function of $x$, $A_1(x)=1-1.25 n_0(x)$. The kinetic energy $K_4$ is proportional to the second moment of $n({\bf k})$. Only $n^{\ast}(k)$ contributes significantly to the second moment so that from (\ref{eqn:mom_distribution}) and with $\bar\alpha_2$ the same in the normal and superfluid phases we have $K_4(S)=A_1(x)K_4(N)$. In this model, the kinetic energy in the superfluid phase (S) is reduced below that in the normal phase (N) entirely by BEC. We found the kinetic energy to be about 10\% lower in the superfluid phase than in the normal phase increasing somewhat with $x$ as shown in Fig. \ref{fig:K4_superfluid}.

\subsubsection{$^3$He momentum distribution}

\begin{figure}[t]
\includegraphics[width=3.3in]{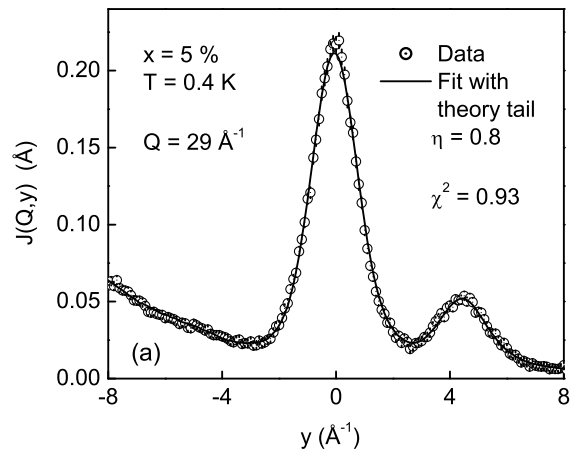}\\
\includegraphics[width=3.3in]{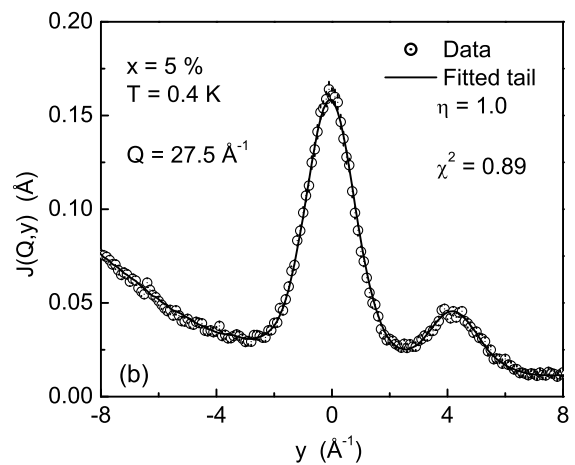}
\caption{\footnotesize Fits to the superfluid helium mixture data at a $^3$He concentration $x=$ 5\%. The $^4$He peak is fitted using the CA method and the $^3$He peak assuming a Fermi $n(\bf{k})$ having a variable tail length $\eta$. The open circles are data points and the solid lines are fits to data. The error bars are the size of the circles except where shown.}
\label{fig:jqylowTfits}
\end{figure}

Fig. \ref{fig:jqylowTfits} shows two fits to the $^3$He peak in the superfluid mixture at $T=$ 0.4 K. In one fit, the tail parameter of the model $n(\bf{k})$ is set at $\eta=$ 0.8, the value that reproduces the calculated tail. The same model ($\eta=$ 0.8) was compared with data for $T=$ 2.5 K in Fig. \ref{fig:jqyfits0}. The agreement with the data for $\eta=$ 0.8 at $T=$ 0.4 K and 2.5 K is indistinguishable. This shows firstly that there is no observable temperature dependence of the $^3$He peak.
between $T=$ 0.4 K and $T=$ 2.5 K which corresponds to $T\simeq T_F$ and $T\simeq 6T_F$, respectively.

 Somewhat surprisingly, there is no observable temperature dependence of the $^3$He momentum distribution although the two temperatures are significantly different. We return to this point in the discussion section. Secondly, there is again no disagreement between theory (for $T=$ 0 K) and experiment. 

As at $T=$ 2.5 K, a range of $\eta$ values provided best fits to data depending upon the $Q$ value considered. Fig. \ref{fig:jqylowTfits} shows an example at $Q=$ 27.5 \AA$^{-1}$ where the best fit is obtained for $\eta=1.0$. At $T=$ 0.4 K, we find $\eta$ lies in the range 0.7$\lesssim \eta \lesssim$ 1.0 or $\eta=$ 0.85$\pm$0.15 as found at $
T=$ 2.5 K.

\subsection{Alternate methods of analysis}\label{sec:resultsC}
In this section, we test the sensitivity of the results to the method of analysis by analyzing the data in different ways.
\subsubsection{Kinetic energy assuming a Gaussian}
\begin{figure}[t]
\centering\epsfig{figure=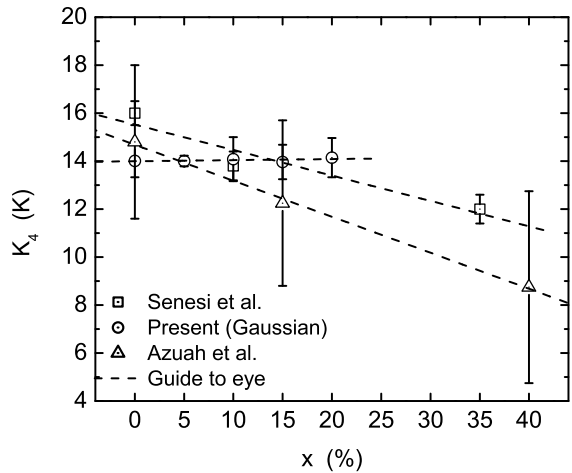,width=3.3in}
\caption{\footnotesize Observed $K_4$ (open circles) assuming a Gaussian fit to the $^4$He $J(Q,y)$. The measured values by Azuah {\it et al.} (open triangles) \cite{Azuah:95} and Senesi {\it et al.} (open squares) \cite{Senesi:03} are shown for comparison. The dashed lines are guides to the eye.}
\label{fig:xdep_reproduce_senesi}
\end{figure}
A straightforward method of analysis is to fit Gaussian functions to both the $^3$He and $^4$He peaks as employed by Azuah {\it et al.}. \cite{Azuah:95} 
This method, when applied to our data yields a $^3$He kinetic energies that agree with those previously obtained \cite{Azuah:95,Senesi:03}  where a Gaussian or a Gaussian-like fit was made. Senesi {\it et al.} \cite{Senesi:03,Senesi} also included the leading term beyond a Gaussian in a Gauss-Hermite expansion of $J(y)$ to analyze their data. However, they found that the inclusion of higher order terms did not result in significant improvements in the fits. Both Azuah {\it et al.} and Senesi {\it et al.} found $K_4$ values that decrease with increasing $^3$He concentration.   

In Fig. \ref{fig:xdep_reproduce_senesi}, we show the $ K_4$ values obtained by fitting a Gaussian $J(Q,y)$ to our data compared with the reported values by Azuah {\it et al.} \cite{Azuah:95} and by Senesi {\it et al.}. \cite{Senesi:03} 
Firstly, for pure $^4$He ($x=$ 0) we find a $K_4$ that lies approximately 10\% below the CA values. A smaller $K_4$ is expected for a Gaussian. In the CA,  FS effects cut off the wings of the peak and allow a somewhat broader $n({\bf k})$ to fit the data. Senesi {\it et al.} took data at $Q\simeq$ 100 \AA$^{-1}$ where FS effects are probably negligible. Thus we expect their values to agree with our CA values. Most importantly, using a Gaussian fit we find a $K_4$ that is independent of $x$. Thus while our CA, the Azuah {\it et al.} and the Senesi {\it et al.} $K_4$ agree for pure $^4$He value ($x=0$), the data itself appears to disagree on the $x$ dependence of $K_4$. Most theoretical calculations also find that $K_4$ decreases with increasing $x$ as shown in  Fig. \ref{fig:k4compared}a. 
\subsubsection{Condensate assuming concentration dependent kinetic energies}
\begin{figure}[t]
\centering\epsfig{figure=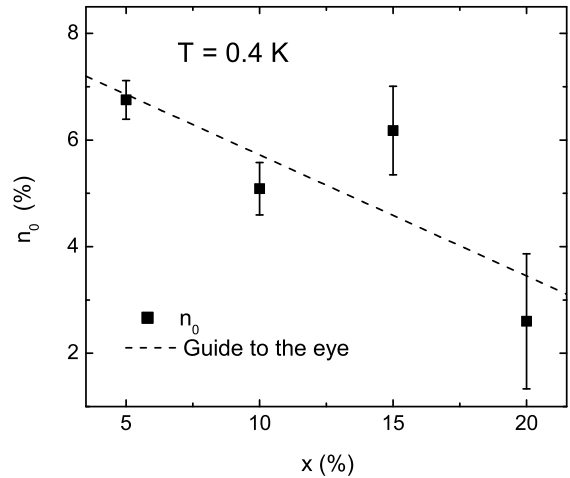,width=3.3in,angle=0}
\caption{\footnotesize Condensate fraction, $n_0$, obtained from present data assuming the kinetic energies measured by Azuah {\it et al.} and Senesi {\it et al.} \cite{Senesi:03}. The $n_{0}$ decreases with increasing $x$. This is illustrates the correlation between $n_0$ and $K_4$.}
\label{fig:K_senesi_n0}
\end{figure}

Since there is some correlation between the $\bar\alpha_2$ parameter ($K_4$) and $n_0$, we determined $n_0$ using the values of $\bar \alpha_2$ corresponding to the measured $^4$He kinetic energies by Senesi {\it et al.}.\cite{Senesi:03} The resulting values of $n_0$ are shown in Fig. \ref{fig:K_senesi_n0}. We find $n_0$ decreases with increasing $x$, when the $x$ dependent $K_4$ values of Senesi {\it et al.} and Azuah {\it et al.} are used. 
This result is expected since a narrower $^4$He peak is obtained by either increasing $n_0$ or decreasing $\bar\alpha_2$. If $\bar\alpha_2$ ($\sim K_4$) decreases, a smaller $n_0$ is needed to obtain a good fit. Thus a $K_4$ that decreases significantly with $x$ implies an $n_0$ that also decreases with $x$ in disagreement with theory.

\section{Discussion and Conclusion}\label{sec:conclusions}

\subsection{\4he response} 

\begin{figure*}
\includegraphics[width=3.3in]{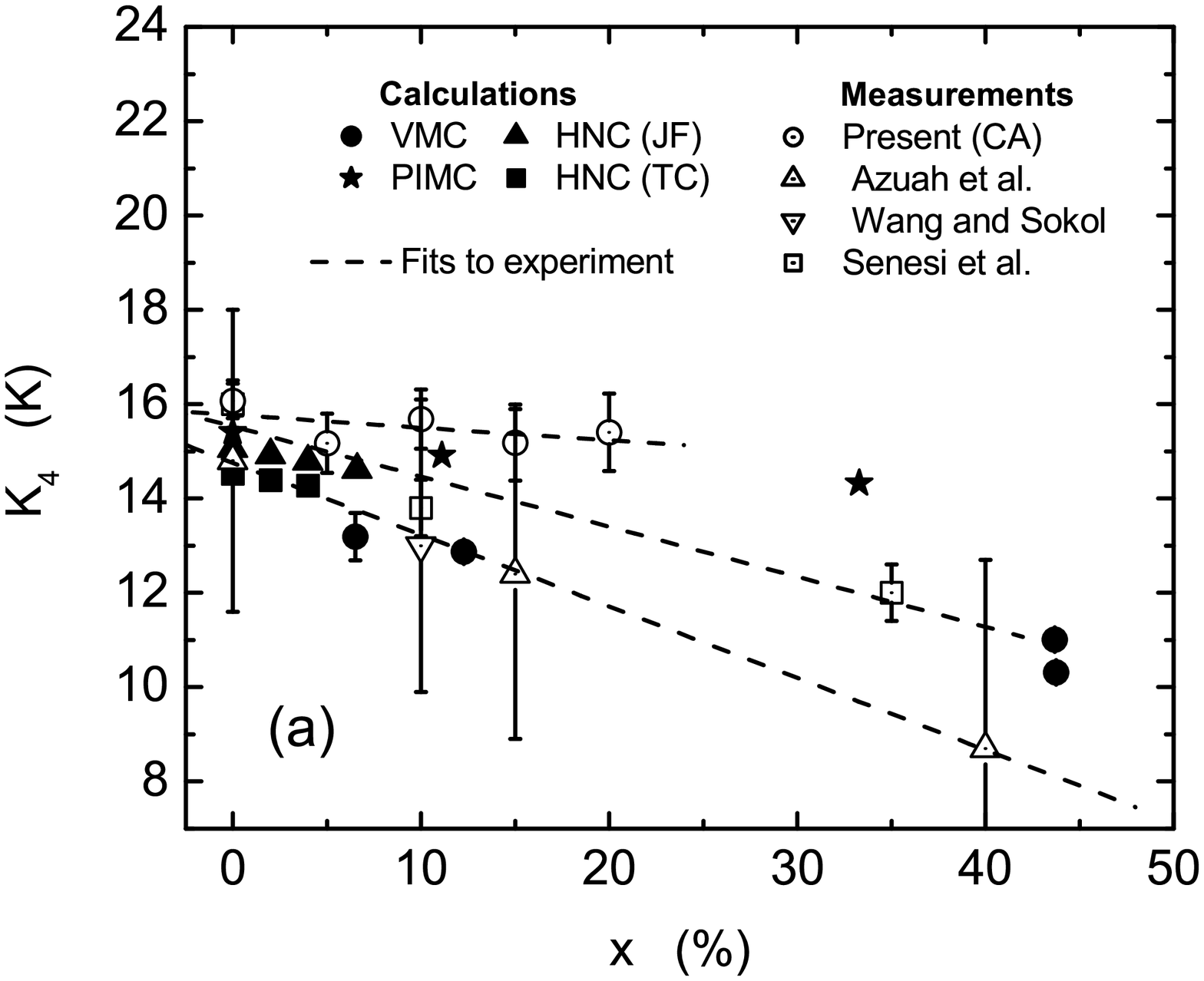}
\includegraphics[width=3.3in]{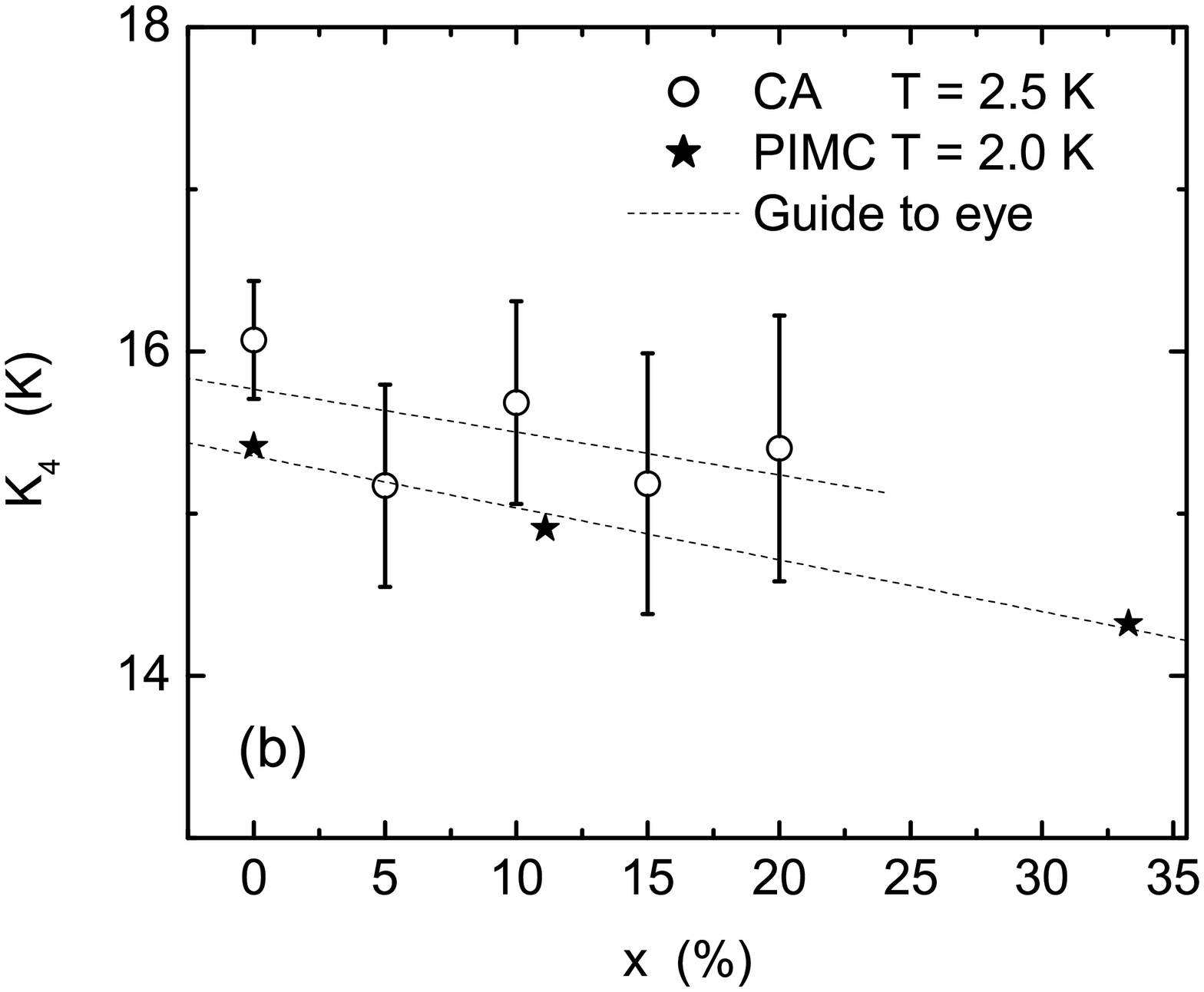}
\caption{\footnotesize {\bf (a)}: $^4$He kinetic energy ($K_4$) as a function of $^3$He concentration, theory and experiment. Calculations are VMC\cite{Lee:81} (solid circles), HNC using a Jastrow function \cite{Boronat:97} (triangles), HNC including triplet correlations \cite{Boronat:97} (solid squares).
Experiments are Azuah {\it et al.}\cite{Azuah:95} (up-pointing triangles), Wang and Sokol \cite{Wang:94} (down-pointing triangle), Senesi {\it et al.}\cite{Senesi:03} (open squares) and present at $T=$ 2.5 K (open circles). Dashed lines are fits to experiments. {\bf (b)}: Present observed $K_4$ at $T=2.5$ K (open circles) as a function of $^3$He concentration compared with a PIMC calculation \cite{Boninsegni:97} of $K_4$  at $T=2.0$ K (solid stars). Dashed lines are guides to eye. 
}
\label{fig:k4compared}
\end{figure*}

In pure normal $^4$He, we find a kinetic energy ($K_4=16.1\pm0.3$ K) which agrees well with previous neutron scattering data \cite{Glyde:00} ($K_4=16.3\pm0.3$ K ) and with calculations \cite{Boninsegni:97} ($K_4=15.41$ K). In normal 
helium mixtures, we find a $K_4$ that decreases somewhat with increasing \3he concentration. This suggests that the local environment of individual $^4$He atoms remains largely unchanged \cite{Dobbs:book} with the addition of $^3$He atoms at low $x$. This finding represents the chief difference between our results and early theoretical treatments \cite{Lee:81,Boronat:97} and previous experiments \cite{Azuah:95,Senesi:03} which find that $K_4$ decreases significantly with increasing $x$. 
Theoretical and experimental values of $K_4$ are compared in Fig. \ref{fig:k4compared}a. The present $K_4$ and a more recent Path Integral Monte Carlo (PIMC) calculation \cite{Boninsegni:97} of $K_4$ which show a substantial agreement on the $x$ dependence are compared separately in Fig. \ref{fig:k4compared}b. Clearly, there remain some theoretical and experimental differences to be resolved. The present $K_4$ in the superfluid phase was obtained from the normal phase value assuming that the condensate does not contribute to the kinetic energy.

To determine the condensate fraction, $n_0$, in helium mixtures, we introduced a model $n({\bf k})$ which has a condensate term and fitted the model to the neutron scattering data. In the model, the second moment of $n({\bf k})$ for the  finite ${\bf k}$ states was assumed to be the same as in the normal phase, as we found previously \cite{Glyde:00} for pure liquid \4he. The higher moments and the Final State function were assumed to be the same as in pure superfluid helium. 

\begin{figure}[t]
\centering\epsfig{figure=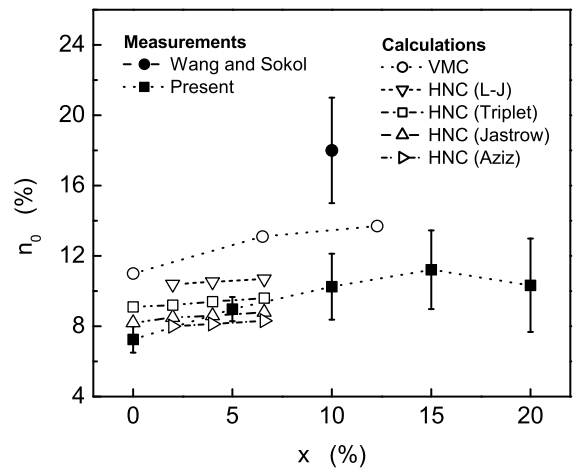,width=3.3in,angle=0}
\caption{\footnotesize The condensate fraction ($n_{0}$) as a function of $^3$He concentration, theory and experiment. Calculations are the open symbols; VMC \cite{Lee:81} (circles), HNC (Lennard-Jones) \cite{Boronat:90} (down-pointing triangles), HNC (triplet correlations) \cite{Boronat:97} (squares), HNC (Jastrow)\cite{Boronat:97} (up-pointing triangles), HNC (Aziz potential) \cite{Boronat:90} (right-pointing triangles). Measurements are Wang and Sokol\cite{Wang:94} (solid circle) and present (solid squares).}
\label{fig:n0_comparison}
\end{figure}

In mixtures we find a small increase in $n_0$ with increasing $^3$He concentration, to $n_0 = 11 \pm 3 \%$ at $x =$ 15-20 \%. Fig. \ref{fig:n0_comparison} shows our observed $n_0$ as a function of $x$ compared theoretical values. Our results agree with the theoretical values within experimental error. Wang and Sokol \cite{Wang:94} measured $n_0$ at one concentration reporting a value of $n_0=$ 18\% at $x=$ 10\%. This is higher than the present measured and calculated values. 
The concentration dependence of $n_0$ may be roughly estimated assuming that the effect of the $^3$He is simply to change the volume available to the $^4$He (average density approximation) and using the density dependence of $n_0$ in pure \4he.\cite{Moroni:04} Essentially, the molar volume of the mixture at low concentrations is approximately \cite{Edwards:69} $v\simeq v_4\left(1+\alpha x\right)$ where $v_4$ is the $^4$He molar volume at SVP and $\alpha$ the excess of volume occupied by $^3$He ($\simeq0.28$). This gives a mixture density which decreases with $x$. On the other hand, the condensate fraction, $n_0$, in pure liquid $^4$He increases \cite{Snow:92,Moroni:04} with decreasing density. This yields an $n_0$ which increases very slightly with $x$, from $n_0\sim$ 7\% ($x=$ 0) to $n_0\sim$ 9\% ($x=$ 20\%). This simple calculation suggests that $n_0$ should increase little with concentration, in qualitative agreement with our experiment.

\subsection{\3he response} 

We find that a model \3he momentum distribution $n({\bf k})$ with a small step, $Z$, at the Fermi surface ($k = k_F$) and a substantial tail at higher $k$ characteristic of a strongly interacting Fermi liquid reproduces the observed \3he recoil peak well. For a gas of non-interacting Fermions at $T =$ 0 K, $n({\bf k})$ is a simple step 
function ($Z=$1); $n({\bf k})=$ 1 for $k < k_F$ and $n({\bf k}) =$ 0 for $k > k_F$. This case is reproduced by our model $n({\bf k})$ in Eq. (\ref{eqn:fermi_nk}) for parameters $H =$ 1, $H_F =$ 0 ($Z= H - H_F =$ 1, no tail in $n({\bf k})$). When Fermions interact, some Fermions move from states below $k_F$ to states above $k_F$ to minimize the total energy, potential 
plus kinetic energy. The step at $k_F$ is reduced $Z <1$ and $n({\bf k})$ develops a tail at higher $k$. This is reproduced in Eq. (\ref{eqn:fermi_nk}) 
for parameters $H - H_F = Z<1$ and $\eta > 0$. The stronger the 
interaction, the smaller is $Z$ and the longer is the tail (smaller $\eta$). In pure liquid $^3$He a recent diffusion Monte Carlo (DMC) calculation 
\cite{Mazzanti:04} finds $Z =$ 0.236 and a tail in $n({\bf k})$ that is well represented by 
an exponential (times algebraic terms) in $k$. This $n({\bf k})$ agrees 
well with earlier DMC results.\cite{Moroni:97,Whitlock:87}
 
Pure liquid $^3$He is regarded as a strongly interacting Fermi liquid, 
more strongly interacting than nuclear matter, and $Z \sim 0.25$ 
represents a large reduction of the step height at $k_F$ arising from 
interaction. As noted above, Boronat {\it et al.} \cite{Boronat:97} calculate an 
$n({\bf k})$ obtaining $Z=$ 0.09 and a tail that is also well fitted with an 
exponential with $\eta =$ 0.8 for a $^3$He-$^4$He 
mixture of $x =$ 6.6 \% $^3$He. Apparently, $^3$He in a dilute mixture is 
more strongly interacting than in pure $^3$He. Our observed, finite temperature $n({\bf k})$ has $Z =0.05 \pm{^{0.14}_{0.01}}$ and $\eta = 0.85 \pm 0.15$ at $x = 5$ \%. It agrees with the calculated $n({\bf k})$ for $T =$ 0 K. $Z =0.05\pm{^{0.14}_{0.01}}$ represents a very small step at $k_F$.

We have not included Final State (FS) effects in our analysis of the \3he recoil peak. The impact of FS effects depends critically on the width and shape of the intrinsic $J_{IA}(y)$ arising from $n({\bf k})$. If $J_{IA}(y)$ is narrow relative to the FS broadening function $R(Q,y)$ or has a narrow component or sharp feature, then the FS effects serve to broaden the narrow component to the width of the FS function. The classic example is the condensate component in superfluid \4he. However, if $J_{IA}(y)$ is broad relative to the FS function, then the FS function has little further broadening impact. For example, the second moment of the FS function is zero. In this event the FS effects can be well represented by a moment expansion \cite{Sears:84,Glyde:94}  in which the leading term is the third moment. This term serves to introduce an asymmetry into $J(Q,y)$ but will not modify its width or whether $J(Q,y)$ has significant tails at larger $y$ arising from tails in $n({\bf k})$.

The FWHM of $R(Q,y)$ at $Q\approx 25$ {\A} is in the range 0.7-1.0 {\A} in pure liquid \3he, in \3he-\4he mixtures and in pure liquid \4he.\cite{Glyde:00,Mazzanti:01,Mazzanti:04} 
The FWHM of $J_{IA}(y)$  of \3he in \3he-\4he mixtures is approximately 2.0 {\A}, largest in dilute \3he mixtures. We expect FS effects firstly to spread out the discontinuity in $J_{IA}(y)$ arising from the step of height $Z$ in $n({\bf k})$. It will be spread out over a width $y\sim$ 0.7 {\A}  (i.e. of order of 2$k_F$). The present instrument resolution function, shown in Fig. \ref{fig:high_vs_low}, which is broader than the FS function at $Q\approx 25$ {\A}, will further spread out the discontinuity. Parenthetically, it is probably largely because of the resolution and FS broadening that we do not observe any apparent broadening of $n({\bf k})$ arising from temperature. Otherwise, the FS function is narrow relative to the width of $J_{IA}(y)$ arising from $n({\bf k})$. Thus we do not expect FS effects to change the overall width of $J(Q,y)$ (i.e. the apparent width of $n({\bf k})$ and whether it has tails or not). However, including FS effects should improve the shape of $J(Q,y)$ as it has been done \cite{Mazzanti:04} in pure liquid \3he. Thus we do not expect FS effect to alter our basic conclusions about $n({\bf k})$.

We note that one cannot use the FS function for pure \3he for mixtures. Essentially, in dilute \3he mixtures the Fermi statistics for only the \3he atoms should be included rather than for all atoms as in pure \3he. We plan to incorporate FS effects in a future publication.

We have strictly used a model $n({\bf k})$ valid for $T = 0$ K since the step $Z$ at $k_F$ is not rounded by thermal effects to fit data taken at $T =$ 0.4 K $\sim T_F$ and $T =$ 2.5 K $\sim 6 T_F$. Stated differently, we obtain an $n({\bf k})$ at two finite temperatures from the data. These \nk are expressed in the form of a $T= 0$ K $n({\bf k})$. The $n({\bf k})$ at the two temperatures are the same within experimental error. 

For a Fermi gas, the thermal broadening of $n({\bf k})$ is well known. At 
$T \sim T_F$ the broadening is significant with $n(k=0)=$ 0.73, $n(k_F)=$ 0.5, $n(2k_F)=$ 0.23. It could be said that there is a rounded step of 
$Z\simeq0.5$ in $n({\bf k})$ over a width of $k \sim 2k_F$ and $n({\bf k})$ develops a short tail. This thermal broadening is less than the resolution and FS broadening. The thermal broadening of $n({\bf k})$ for an interacting 
Fermi liquid is not known. However, in a strongly interacting Fermi liquid 
$Z$ is already small and $n({\bf k})$ already has a long tail. Hence 
thermal broadening is expected to be relatively less important in an interacting 
liquid. For example, a higher temperature might be required before the 
thermal rounding (e.g. effective thermal reduction of $Z$) becomes significant 
relative to the \lq interaction' reduction. Apparently, this rounding is not 
significant or critical compared to the rounding arising from FS effects 
or the instrumental broadening within current error. The model parameters $Z = 0.05 \pm{^{0.14}_{0.01}} $ and $\eta = 0.85 \pm 0.15$ obtained are consistent with theory for $T = 0 $ K.  Stated differently, the calculated $n({\bf k})$ for $T =$ 0 K 
reproduces our experimental data. Our observed $Z$ may be somewhat smaller than the calculated value because of resolution and thermal effects. The theory and experiment are therefore consistent in finding that $^3$He in dilute $^3$He-$^4$He mixtures is a strongly interacting Fermi liquid. This strong interaction must arise at least in part from interaction with the $^4$He component of the liquid in dilute mixtures.

Mazzanti and co-workers \cite{Mazzanti:04} have compared their calculated $J(Q,y)$ for pure liquid $^3$He with that observed by Azuah {\it et al.} at $T=$ 1.4 K ($T \sim T_F$) and $Q = 19.4$ {\AA$^{-1}$} in much the same way we have done here. The input to their calculated $J(Q,y)$ is $n({\bf k})$ for $T=0$ K calculated using DMC. This yields $Z =$ 0.236 as noted above. They 
represented their calculated $n({\bf k})$ by an analytic function that had 
an exponential tail times algebraic terms. The second input is a Final State 
broadening function valid for a pure Fermi liquid. The Final State 
broadening function $R(Q,y)$ has a FWHM in $y$ of approximately 0.7 {\A} 
compared with the total FWHM of $J(Q,y)$ of 2.0 {\A}   so that $R(Q,y)$ 
has only a modest impact on $J(Q,y)$. They find their $J(Q,y)$ reproduces the observed $J(Q,y)$ well and conclude that here is agreement between theory and experiment. In a similar way we have used a model 
$n({\bf k})$ with parameters (rather than a calculated $n({\bf k})$) 
represented by an analytic function with an exponential tail. We find a 
$J(Q,y)$ (without FS effects) which reproduces experiment within 
statistical error. The model $n({\bf k})$ is consistent with \nk 
calculated for mixtures.\cite{Boronat:97}
 We conclude in a similar way that there is no disagreement between theory and experiment for mixtures as well as in pure $^3$He.

In summary, we have measured the momentum distribution 
of ${^3}$He-${^4}$He mixtures for
$^3$He concentrations $x$ between 0 and 20\%. We find $n_0$ increases from 
7.25$\pm$0.75\% in pure $^4$He ($x=$ 0\%) to 11$\pm$3\% for mixtures with 
$x= 15-20\% $, in agreement with theoretical calculations. The $^4$He 
kinetic energy is found to decrease slightly with $^3$He concentration. The model \3he $n(\bf{k})$ that reproduces our data at $x = 5 \%$ agrees with the calculated $n(\bf{k})$ at $x = 6.6 \%$ removing a previous apparent disagreement between theory and experiment based on comparing \3he kinetic energies.

\section*{Acknowledgments}
We thank R.B.E. Down for his assistance on the beamline and J.W. Taylor for technical assistance with the MARI spectrometer. Support of this work by the US department of energy grant DE-FGOZ-03ER46038 and beamtime at the CCLRC ISIS facility is gratefully acknowledged.

\end{document}